\documentstyle [12pt]{article}
\makeatletter
\@addtoreset{equation}{section}
\makeatother

\newcommand{\mathrm}{\rm}
\newcommand{\ba}{\begin{eqnarray}}
\newcommand{\ea}{\end{eqnarray}}

\newcommand{\nll}{\nonumber \\}
\begin{document}
\newcommand{\BS}{\bigskip}
\newcommand{\SECTION}[1]{\BS{\large\section{\bf #1}}}
\newcommand{\SUBSECTION}[1]{\BS{\large\subsection{\bf #1}}}
\newcommand{\SUBSUBSECTION}[1]{\BS{\large\subsubsection{\bf #1}}}
%
%
%
\begin{titlepage}
\begin{flushright}
{UGVA-DPNC 1995/6-166}
\\
{DESY 95-100}
\\
June 1995
\end{flushright}
\begin{center}
\vspace*{2cm}
{\large \bf
BHAGENE3, A MONTE CARLO EVENT GENERATOR
FOR LEPTON PAIR PRODUCTION AND
WIDE ANGLE BHABHA SCATTERING IN $e^{+} e^{-}$
COLLISIONS NEAR THE Z-PEAK
}
\vspace*{1.5cm}
\end{center}
\begin{center}
{\bf J.H.Field \footnote{ E-mail JFIELD@CERNVM.CERN.CH}}
\end{center}
\begin{center}
{
D\'{e}partement de Physique Nucl\'{e}aire et Corpusculaire
 Universit\'{e} de Gen\`{e}ve . 24, quai Ernest-Ansermet
 CH-1211 Gen\`{e}ve 4.
}
\end{center}
\begin{center}
{\bf T.Riemann \footnote{ E-mail riemann@ifh.de}}
\end{center}
\begin{center}
{Deutsches Elektronen-Synchrotron DESY
\\
Institut f\"ur Hochenergiephysik, Platanenallee 6, D--15738 Zeuthen, Germany
}
\end{center}
\vspace*{2cm}
\begin{abstract}
A new Monte Carlo event generator for wide angle Bhabha scattering
and muon pair production in $e^+e^-$ collisions is described. The
program includes complete one-loop electroweak corrections, and QED
radiative corrections. The O($\alpha$) QED correction uses the exact
matrix element. Higher order QED corrections are included in an
improved soft photon approximation with exponentiation of initial
state radiation. Events are generated in the full phase space of the
final state including explicit mass effects in the region of collinear
mass singularities. The program is intended for centre of mass energies
around and above the Z peak and for Bhabha scattering at angles greater than
$10^{\circ}$ .
\end{abstract}
\vspace*{1cm}
\end{titlepage}
\begin{center}
{\bf PROGRAM SUMMARY}
\end{center}
\par \underline{\it Title of the program: } BHAGENE3
\par \underline{\it Computer: } IBM 3090
\par \underline{\it Operating system:} VM/CMS
\par \underline{\it Programming language used:} FORTRAN 77
\par \underline{\it High speed storage required:} 678k words
\par \underline{\it No. of bits in a word:} 32
\par \underline{\it Peripherals used:} Line printer
\par \underline{\it Number of cards in combined program and test deck:}
 about 6900
\par \underline{\it Nature of physical problem:} Calculation of all 1-loop
electroweak corrections, and of QED corrections up to O($\alpha^3$) for wide
angle Bhabha scattering and muon pair production in the vicinity of the Z peak.
\par \underline{\it Method of solution:} The 1-loop electroweak virtual
corrections are calculated analytically. The O($\alpha$) QED radiative
corrections use the
exact matrix element. The higher order QED corrections are given by an improved
 soft photon approximation. Event configurations containing real photons are
 first generated according to an approximate model, and are then reweighed
 according to the theoretical distributions. The weight throwing technique is
 used to produce unit weight events in the full final state phase space.
\par \underline{\it Restrictions on the complexity of the problem: } Some
terms in $m_l$/$E$ ($l=e,\mu$) are neglected, so that fermion threshold effects
are not properly taken into account. The approximation used for O($\alpha^2$)
and higher order real photon radiation may be unreliable in events with
multiple
hard photon radiation. For cross section accuracy at the \% level it is not
recommended to use the program for Bhabha scattering angles less than
$10^{\circ}$.
\par \underline{\it Typical running times:} For Bhabha events: initialisation
 about 140 sec, generation 709 unit weight events per second. For muon pairs
 the initialisation time is shorter by a factor of about 3 and the event
 generation rate is roughly doubled. The time unit is an IBM3090 CPU second.
\par \underline{\it Unusual features of the program:} Extensive use is made of
one and two dimensional look-up tables for fast and flexible generation of
Monte-Carlo variables. The space required for these tables means that the fast
memory requirements may be greater than in other comparable programs. These
look-up tables as well as average event weights are created in the
initialisation
 phase of the program which is, in consequence, relatively time-consuming.
\par \underline{\it References:}
\begin{itemize}
\item[[1]] J.H.Field, Phys. Lett. B323 (1994) 423
\item[[2]] M.B\"{o}hm, A.Denner and W.Hollik,
Nucl. Phys. B304 (1988) 687.
\item[[3]]  F.A.Berends, R.Kleiss and W.Hollik,
Nucl. Phys. B304 (1988) 712.
\item[[4]] D.Bardin, W.Hollik and T.Riemann,
Z. Phys. C Particles and Fields 49 (1991) 485.
\end{itemize}

\SECTION{\bf{Introduction}}

Data on the leptonic decays of the Z taken since 1990 at LEP and SLC
have made possible many precise tests of the Standard Model (SM) of
weak and electromagnetic interactions~\cite{x1}. These electroweak
analyses require Monte Carlo event generators, incorporating both the
SM predictions and the numerically important pure QED radiative
corrections, in order to properly take into account detector acceptance,
efficiency and resolution effects, for arbitrary experimental cuts.

For fermion pair production (excluding $e^{+}e^{-}$), the most widely
used generator is KORALZ~\cite{x2}. This incorporates one-loop
electroweak corrections using the SM, multiple bremsstrahlung photons in
the initial state and a single bremsstrahlung photon in the final state
as well as a complete simulation of $\tau$ decays, including polarisation
effects and QED radiative corrections associated with the $\tau$ decay
products.

For wide angle Bhabha scattering two generators have been written prior
to that, BHAGENE3, described in the present paper. The first, BABAMC
{}~\cite{x3,x4} includes one-loop SM electroweak corrections and QED
corrections
(real and virtual) to O($\alpha$). The second, BHAGEN~\cite{x5}
includes higher order QED corrections (by exponentiation) in both initial
and final states, but is limited to `quasi-elastic' configurations where
the photons are soft, or almost collinear with the radiating lepton.
This limitation is a severe disadvantage when comparing with actual
experimental data, since in general very loose cuts (typically none at
all on the lepton-photon angles) are made in order to minimise the size
of the QED radiative corrections. More recently a new Monte Carlo
generator
UNIBAB~\cite{x6} has been written, which is a general purpose program
in many ways comparable to BHAGENE3.
 Electroweak corrections are included and arbitrary numbers of initial
 and final state photons generated using a LLA structure function
 formalism. Unlike BHAGENE3 however, O($\alpha$) initial/final
 interference effects are not included in the current version.

Bhabha generators also exist ~\cite{x7,x8} for the small angle region of
lepton scattering angle $\theta_{l}$ (dominated by the $t$-channel photon
exchange diagram) typical of the angular acceptance of the luminosity
monitors of the LEP/SLC experiments.

The generator described in this paper, BHAGENE3, is adapted to the
description of wide angle Bhabha scattering ($10^{\circ}  <  \theta_{l}
< 170^{\circ}$), and of $\mu$-pair production, in the region of the Z
peak. $\tau$-decays are not incorporated. Electroweak corrections in the
SM, including the most important two-loop effects, are implemented as
described in Ref.\cite{x9}. Multiple hard photon generation, ( $n_{\gamma}
\leq 2(3)$ for initial (final) states) is also included. No restrictions
are necessary on the kinematical configuration of the generated events.
However, the approximate nature of the hard photon generation algorithm
should be borne in mind when multiple, very hard photon configurations
are considered~\cite{x10}. A brief description of the program has been
given previously, and comparisons of cross-sections and charge
asymmetries made~\cite{x10} with the analytical or semi-
analytical programs:
 ZFITTER~\cite{x11}, ALIBABA~\cite{x12} and TOPAZ0~\cite{x13}.

The approach adopted for the QED radiative corrections is to treat the
O($\alpha$) correction exactly (in particular the relevant fermion mass
terms are systematically included throughout, so that collinear photon
radiation is correctly generated) as in BABAMC~\cite{x3,x4} and
MUSTRAAL~\cite{x14}. Higher order real photon radiation is treated, not
by using exact O($\alpha^{2}$) amplitudes~\cite{x15,x16,x17}, but by an
improved soft photon approximation. This may be justified~\cite{x10} by
the relatively small size of the O($\alpha^{2}$),O($\alpha^{3}$), ...
hard photon corrections as compared to that at O($\alpha$). As in
Ref.[3,4,14] the final state phase space is divided into two regions.
A cut ($E^{\gamma}_{0}$) is applied on the energy of each photon. For
$ E^{\gamma} < E^{\gamma}_{0}$ the photon energy and angular
distributions are integrated over, so as to yield a Virtual, Soft (VS)
corrected cross-section. The corresponding generated events have a
`Born topology'; the outgoing leptons being exactly back-to-back with
energy: $ E_{l} = E_{beam} = E $. The generation of such V,S events
is described in Section 3 below. For $ E^{\gamma} > E^{\gamma}_{0} $ a
dilepton event with 1,2,3 hard photons is generated over the full
available phase-space. Denoting the number of initial/final state
photons by $n^{I}_{\gamma}/n^{F}_{\gamma}$, the different possibilities
are:
\[ n^{I}_{\gamma}/n^{F}_{\gamma} = 1/0, 0/1, 2/0, 1/1, 0/2, 0/3 \]
This `hard photon' generation is described in Section 4 below.

A complete description of the program, its structure and parameters,
may be found below in Section 5. In this Introduction only a brief
account is given of the most important techniques employed (see
also Ref.[10]). The first step in the execution of the program is an
initialisation phase where all electroweak parameters are calculated
within the SM from standard input parameters such as the fine
structure constant, $\alpha$, the Fermi constant G, and the masses of
the Z ($M_{Z}$), the top quark ($M_{t}$) and the Higgs boson ($M_{H}$).
Parameters needed for the QED radiative corrections, taking into account
the lepton flavour (e,$\mu$) selected and the beam energy E are also
calculated at this stage. Next the probabilities for VS, `Initial
State hard' (IS) and `Final State hard' (FS) events are calculated.
For the VS events this is done by angular integration of the
differential cross-section, including weak and electromagnetic
radiative corrections. For IS, FS events approximate factorisable
cross-sections are used that may be readily integrated analytically
and/or numerically, to give the corresponding probabilities. Also,
during the initialisation phase, a number of one or two dimensional
Look Up Tables (LUT) for the total photon energy, photon energy
splitting fractions, and lepton scattering angle, are produced.
For each variable the Integrated Probability Distribution (IPD)
is calculated by analytical and/or numerical integration.
The
IPD is then inverted by linear interpolation (see Appendix C) to
yield the LUT. The average weights $\overline{W}_{I}$, $\overline{W}_{F}$
of the IS, IF events are also found, during initialisation, by
re-weighting events according to the `exact' hard cross-section,
to be described below in Section 4.

The event generation phase of the program execution is now entered.
A
VS , IS or FS event is first chosen. For IS, FS events $n^{I}_{\gamma}$
or $n^{F}_{\gamma}$ are then chosen according to a Poisson distribution,
and the appropriate sub-generator is entered. The extensive use of LUT
rather than a Weight Rejection Procedure (WRP) as in Ref.[4,13] gives
a very fast and efficient event generation algorithm. Arbitrary user
defined cuts may be applied during the event generation phase.
Four-vectors
are written out for each unit weight event which is chosen by
using a WRP.  In the last stage of execution the
`exact' cross section corresponding to the user-supplied cuts is is
 calculated and printed out as well as the user-defined histograms
 or other distributions that are up-dated during the generation phase.

Finally in this Introduction a few remarks on the limitations of the
program and some recommended restrictions on its use. Only cross-sections
 summed over final and averaged over initial states of the
lepton and photon polarisations are calculated. Only leptonic (not
quark anti-quark) final states may be generated. As in Ref.[3,4,14]
not all terms $ \approx m_{l} / E $ are included, so that the
near-threshold region of lepton pair production is not properly
described. As an approximate model is used to generate hard photons
at O($\alpha^{2}$) and higher, based on an improved soft photon
approximation, cross-sections for configurations with multiple
hard photons should be regarded with  caution. Some comparisons
with exact O($\alpha^{2}$) calculations are given in Ref.[10].
The precision of the LUT's used for the lepton scattering angle
requires that $10^{\circ} < \theta_{l} < 170^{\circ}$ in Bhabha
scattering, if a cross-section accuracy of $\approx 1\%$ is required.
%
%
\SECTION{\bf{Weak Virtual Corrections}}
%
In the initialisation, the $W$ mass is determined iteratively from $\alpha,
G_{\mu}, M_Z$ in subroutine SETCON:
\ba
M_W &=& M_Z \sqrt{1-\sqrt{1-\frac{4\pi\alpha}
{\sqrt{2}G_{\mu}M_Z^2(1-\Delta r)}}}
\label{mwiter}
\ea
where $\Delta r$ is calculated in subroutine SEARC1.

The improved Born approximation for the Bhabha differential cross section
 consists of the
sum of contributions with $t$ channel and $s$ channel exchange of the
photon and $Z$ boson and their interferences:
\ba
\frac{d \sigma^{WC}}{dc}
=
\frac{\pi \alpha^2}{2 s} \sum_{A} \sum_{a} T_a(A)
& A=\gamma, \gamma Z, Z,
& a= s,st,t
\label{v1}
\ea

 Here $s$ and $t$ are the usual Mandelstam variables and $c = \cos \theta_l $
 where $\theta_l$ is the lepton scattering angle.
In Bhabha scattering the final state helicities are not measurable, while
the beams may be longitudinally polarized.
We introduce the notation:
\ba
\lambda_1 &=& 1-\lambda_+ \lambda_-
\\
\lambda_2 &=& \lambda_+ - \lambda_-
\\
\lambda_3 &=& 1+\lambda_+ \lambda_-
\label{lamda}
\ea
where $\lambda_{+(-)}$ is the degree of longitudinal polarisation
 of the positron (electron) beam.

The s channel cross section contributions are:
\ba
 T_s(\gamma)
&=&
\left| F_A(s)\right|^2
\left[ \lambda_1 \left( 1+c^2 \right) \right]
\\  
T_s(\gamma Z)
&=&
2 \Re e
\Bigl\{
\rho(s) \chi(s)
F_A^*(s)
\Bigl[
[ \lambda_1 v_{ee}(s) + \lambda_2 v_e(s) ] (1+c^2)
\nll 
&&
+~ [(\lambda_1+\lambda_2) v_e(s) ] 2c
\Bigr] \Bigr\}
\\ 
T_s(Z)
&=&
\left|\rho(s) \chi(s) \right|^2
\Biggl\{
\Bigl[
\lambda_1 \left( 1 + 2 |v_e(s)|^2 + |v_{ee}(s)|^2 \right)
\nll 
& & 
+~2 \lambda_2 \Re e \left( v_e(s) [1+v_{ee}(s)^*]\right) \Bigr] (1+c^2)
\nll
&&
+~2 \Re e
\left[
\lambda_1\left(|v_e(s)|^2 + v_{ee}(s)\right)+\lambda_2 v_e(s)
\left( 1+v_{ee}^*(s) \right)
\right]
2c
\Biggr\}
\label{v2}
\ea
Here, we use the following abbreviations:
\ba
\chi(s) &=& \kappa \frac{s}{s-M_Z^2 + is\Gamma_Z / M_Z}
\label{chiz}
\\
\kappa &=& \frac{G_{\mu}}{\sqrt{2}} \, \frac{M_Z^2}{8\pi \alpha}
\\
v_e(s) &=& 1 - 4 s_W^2 \kappa_e(s)
\\
v_{ee}(s) &=& -1 + 2 v_e(s) + 16 s_W^4 \kappa_{ee}(s)
\label{v3}
\ea
In our expression~(\ref{v1}) for $d\sigma^{WC}/dc$
 the axial couplings are equal to unity.
The effective
couplings $v_e(s), v_{ee}(s)$ and the weak form factor $\rho(s)$ are complex
valued.
They contain virtual weak
corrections which are called from the electroweak
library {\tt BHASHA} which is part of the program package described
here.
It has been derived from the electroweak library {\tt DIZET}~\cite{x18}
which is used in the package {\tt ZFITTER}~\cite{x11} and is
intended for the description of $s$ channel fermion pair production in the
region of the Z resonance.
The complete theoretical description of the renormalisation procedure adopted
in the unitary gauge and related topics may be found in~\cite{x19},
the formulae for the $Z$ width in~\cite{x20}  and $s$ channel scattering
in~\cite{x11}, and those for Bhabha scattering in~\cite{x9}.
The complex valued
$s$ channel corrections from the fermionic vacuum polarisation are contained
in $F_A(s)$,
\ba
F_A(s)
&=& \frac{\alpha(-|s|)}{\alpha}
\label{fas}
\ea
and are explained below.

We now describe the t channel contributions:
\ba
T_t(\gamma)
&=&
F_A(t)^2
\left[
2 \lambda_1
\frac{(1+c)^2}{(1-c)^2}
+ 8 \lambda_3
\frac{1}{(1-c)^2}
\right]
\\  
T_t(\gamma Z)
&=&
2 \rho(t) \chi(t) F_A(t)
\nll
&& \times \left\{
2 \left[ \lambda_1 \left( 1+v_{ee}(t) \right) - \lambda_2 v_e(t) \right]
\frac{(1+c)^2}{(1-c)^2}
- 8 \lambda_3 \left(1-v_{ee}(t)\right)
\frac{1}{(1-c)^2}
\right\}
\nll
\\  
T_t(Z)
&=&
\left[\rho(t) \chi(t) \right]^2
\Biggl\{
2\Biggl[
\lambda_1 \left( 1+4v_e(t)^2+2v_{ee}(t) +v_{ee}(t)^2\right)
\nll
& &
+4\lambda_2 v_e(t) \left(1+v_{ee}(t)\right) \Biggr]
\frac{(1+c)^2}{(1-c)^2}
+~8\lambda_3\left(1-v_{ee}(t) \right)
\frac{1}{(1-c)^2}
\Biggr\}
\label{v5}
\ea
The following additional abbreviations are used:
\ba
\chi(t) &=& \kappa \frac{t}{t-M_Z^2}
\\
t&=& -\frac{s}{2}\left(1-c\right)
\label{v6}
\ea
The form factors $F_A(t), v_e(t), v_{ee}(t), \rho(t)$ are real valued.
For wide angle Bhabha scattering, the values of $|t|$ may become substantially
smaller than $s$.
Nevertheless, the form factors do not vary much since they depend only
logarithmically on the scale.
There is one exception to this.
Bremsstrahlung diagrams with initial state radiation
and $t$ channel photon exchange may yield substantial cross section
contributions.
There, the effective value $|t'|$ of $|t|$
may become extremely small and the different value of the running QED coupling
has to be taken into account properly:
\ba
F_A(t)
&=& \frac{\alpha(|t|)}{\alpha}
\label{fat}
\ea
 For this reason the running alpha correction is included not only in
 $d\sigma^{WC}/dc$ but also in the cross sections for final states with
 hard photons (see below).

Finally, we describe in this section the contributions from the $\gamma
Z$ interference:
\ba
T_{st}(\gamma)
&=&
- 2 \Re e \left[ F_A^*(s) F_A(t) \right] \lambda_1
\frac{(1+c)^2}{(1-c)}
\\  
T_{st}(\gamma Z)
&=&
- 2 \Re e
\left\{
\chi(t) \rho(t) F_A^*(s)
\left[ \lambda_1 \left( 1+v_{ee}(t) \right) - 2 \lambda_2 v_e(t) \right]
+ \left( t \leftrightarrow s \right)
\right\}
\frac{(1+c)^2}{(1-c)}
\nll
\\  
T_{st}(Z)
&=&
-2 \Re e \left\{ \chi(s) \rho(s) \chi(t) \rho(t) \left[
\lambda_1 \left( [1+v_{ee}(s)]  [1+v_{ee}(t)]
\right. \right. \right.
\nll
& &
\left. \left. \left.
+ 4 v_e(s) v_e(t) \right)
- \lambda_2 \left( v_e(s) [1+v_{ee}(t)] + v_e(t) [1+v_{ee}(s)] \right)
\right] \right\}
\frac{(1+c)^2}{(1-c)}
\nll
\label{v7}
\ea

The running of the QED coupling~$\alpha(Q^2)$
is taken into account as follows:
\ba
\alpha(Q^2)
&=&
\frac{\alpha}
{1-\Delta \alpha}
\\
\Delta \alpha &=&
\Delta \alpha_l + \Delta \alpha_{udcsb} + \Delta \alpha_t
\label{dalf}
\ea
where we use the convention that in the $s$ channel it is $Q^2=-s$, and in the
$t$ channel $Q^2=|t|$.
The $\Delta \alpha$ is calculated in function XFOTF1.

For leptons:
\ba
\Delta \alpha_l &=& \sum_{f=e,\mu,\tau}
Q_f^2 N_f
\Delta F_f(Q^2)
\label{vac0}
\\
\Delta F_{f}(Q^2)
&=&
\frac{\alpha}{\pi}
\left\{-\frac{5}{9}+\frac{4}{3}\frac{m_f^2}{Q^2}+\frac{1}{3}
\beta_f
\left(1-\frac{2m_f^2}{Q^2}\right)
\left[
\ln \left| \frac{\beta_f+1}{\beta_f-1} \right| - i \pi \theta(-Q^2-4m_f^2)
\right] \right\}
\label{vac1}
\nll
\\
\beta_f
&=&
\sqrt{1+\frac{4m_f^2}{Q^2}}
\label{vac2}
\ea
and their colour factor $N_f$ and charge $Q_f$ are unity.
In the weak library, the $\Delta F_f$ is calculated by function XI3,
$\Delta F_f = 2$ XI3.

The contribution from the light quarks has been parametrised in two different
ways.
Either it is calculated with~(\ref{vac0})
(with flag setting NPAR(2)=2); in this case,
we use effective quark masses~\cite{x21}:
$m_u=m_d=0.041, m_s=0.15, m_c=1.5, m_b=4.5$ GeV.
The other, preferred approach (with NPAR(2)=3) uses a parametrisation of the
hadronic vacuum polarisation~\cite{x19}, which is contained in function XADRQQ.

The $t$ quark corrections are:
\ba
\Delta \alpha_t &=&
Q_t^2 N_t
\Delta F_t(Q^2) + \Delta \alpha
^{\mathrm{2loop},\alpha\alpha_s}
\label{vact}
\ea
where the latter term contains higher order corrections and is calculated from
functions ALQCDS, ALQCD:
\ba
\Delta \alpha
^{\mathrm{2loop},\alpha\alpha_s}
&=&
\frac{\alpha\alpha_s}{3\pi^2}Q_t^2\frac{m_t^2}{Q^2}
{\cal R}e
\Biggl\{
\Pi_t^{VF}(Q^2) + \frac{45}{4}
\Biggr\}
\label{daqcd}
\ea
 $\Pi_t^{VF}(Q^2)$ is a two loop self energy
function~\cite{x22,x23}.
Setting the corresponding flag NPAR(3) to zero (default value) results in the
neglect of this very small correction. For NPAR(3)=1,2 approximate, exact
calculations according to Eqn.(2.29) are made.
 The first one is not a really good approximation, the second is
very time consuming.

For the function $\Delta F_f(Q^2)$ one may derive the following two
approximations.
For light fermions
 with $m_f^2 << |Q^2|$, the following approximate formula is valid:
\ba
\Delta F_f(Q^2)
\rightarrow
\frac{\alpha}{3\pi}
\left[ \ln\frac{|Q^2|}{m_f^2}-\frac{5}{3}-i\pi\theta(-Q^2)\right]
\label{runalf}
\ea

While, heavy fermions with $m_f^2 >> |Q^2|$
practically decouple:
\ba
\Delta F_f(Q^2)
\rightarrow
\frac{\alpha}{3\pi}
\left( \frac{4}{15}\frac{-Q^2}{m_f^2}\right)
\label{runalt}
\ea

At LEP~1, the effective QED coupling may be treated as a constant in the $s$
channel~\cite{x24}:
\ba
F_A(M_Z^2) \approx \frac{137.036}{128.87}
\label{v4}
\ea

The form factors $\kappa_{e}, \kappa_{ee}, \rho$ are calculated
in subroutine
ROKAP($\ldots$, s, t, u,$\ldots$, QE, QF, $\ldots$, XFF,$\ldots$).
XFF is a vector of 4 functions\footnote{
The third of these four functions, $v_f$, is for Bhabha scattering equal to the
second one, $v_e$.
}
which depend on $s$,$t$,$u$ and the charges of the
initial and final state fermions (here minus one):
\ba
u = -t -s
\label{u}
\ea
and
\ba
\rho(s)  &=& \mbox{XFF}(1; u, -s, t; -1,-1)
\nll
v_e(s)   &=& \mbox{XFF}(2; u, -s, t; -1,-1)
\nll
v_{ee}(s)&=& \mbox{XFF}(4; u, -s, t; -1,-1)
\\
\rho(t)  &=& \mbox{XFF}(1; s,  -t,  u; -1,-1)
\nll
v_e(t)   &=& \mbox{XFF}(2; s,  -t,  u; -1,-1)
\nll
v_{ee}(t)&=& \mbox{XFF}(4; s,  -t,  u; -1,-1)
\label{v8}
\ea
These corrections are switched on and off with flag NPAR(1);
the order $\alpha^2m_t^4$ contributions to them with NPAR(6),
the $\alpha\alpha_s m_t^2$ corrections with flag NPAR(3), and
the $ZZ$ and $WW$ box terms with NPAR(4).
The latter are negligibly small at LEP~1\footnote{
This statement depends to a certain extent on the calculational scheme
chosen.}.
They introduce in the weak
corrections to the $s$ channel a dependence on the scattering angle.
In the $t$ channel, correspondingly,
 the weak corrections will depend not only on the scattering
angle, but
also on s.

The total Z width~\cite{x20} is used in~(\ref{chiz}) and
is calculated in subroutine ZWRATE:
\ba
\Gamma_Z
&=&
\frac{G_{\mu}M_Z^3}{12\pi\sqrt{2}} \sum_f N_C(f) R_f^{QCD}
\sqrt{1-\frac{4m_f^2}{M_Z^2}}
\left(1+\frac{3}{4}\frac{\alpha}{\pi}Q_f^2\right)
\rho_f^Z
\nll
&&\times~
\left\{
\left(1+\frac{2m_f^2}{M_Z^2}\right)
\left[1-4 \kappa_f^Z s_W^2 |Q_f| +8 \left(\kappa_f^Z\right)^2 s_W^4 Q_f^2
\right]
-\frac{3m_f^2}{M_Z^2}
\right\}
\label{gammaz}
\ea
The form factors $\rho_f^Z, \kappa_f^Z$ are not identical with the process
dependent form factors for the scattering process although the leading terms of
the latter, if calculated at the $Z$ peak, yield a good approximation of the
first ones.
The overall factor $N_C(f)$ is equal to 1 for leptons and 3 for quarks and
$R_f^{QCD}$ describes final state QCD corrections in case of quarks:
$R_f^{QCD} = 1+\alpha_s/\pi + \ldots$.
The numerical value has to be chosen in accordance with the
scheme and order of the QCD corrections in mind and is set by the user
with the parameters XPAR(10) for the $b$ quark and XPAR(9) for the other, light
quarks.
More details on the $Z$ decay rate may be found in~\cite{x11,x25} and
references therein.
\SECTION{\bf{Virtual and Soft Photonic corrections}}

Three distinct contributions to the virtual and soft radiative
correction may be distinguished:

1) Purely weak virtual corrections
as described in the preceeding Section. These include all self-energy and
vertex corrections involving W, Z and H (Higgs) bosons, as well as ZZ
and WW box diagrams.

2) Electromagnetic virtual corrections: vertex corrections involving
only photons, $\gamma \gamma$ and $\gamma$Z box diagrams and the effect
of both leptonic and hadronic Vacuum Polarisation Insertions (VPI) in all
off-shell photon propagators.

3) Corrections due to real soft photons.

For 1) BHAGENE3 uses the `Dubna-Zeuthen' (DZ) renormalisation scheme
described in Refs.\cite{x19} and [9]. The weak virtual corrections modify the
 vector and
axial-vector coupling constants appearing in the $s$, $t$ channel Z-exchange
diagrams according to Eqns. 3.15-3.18 of Ref.[9]. For 2), the O($\alpha$)
vertex and $\gamma \gamma$ and $\gamma$Z box contributions are taken from
Ref.[3], and the VPI contribution from Ref.[27]. Vertex corrections to
O($\alpha^{2}$) and the corresponding exponentiated soft photon
correction, 3), (the integral over the angles and energies of all photons
with total energy below a fixed cut-off: $ y = E^{\gamma}_{tot}/E < y_{0}
$) are given by Ref.[28,29]. The differential cross-section, corrected
for weak loop effects 1), and also the VPI part of 2), is denoted by
$ d^{WC}\sigma/dc $ where $c = \cos \theta_{l}$ , and $\theta_{l}$ is the
lepton scattering angle. Including the soft real photon, the O($\alpha$)
virtual, and the leading log O($\alpha^{2}$) virtual corrections as well
as the $\gamma \gamma$, $\gamma$Z box diagrams, gives the following
expression for the Virtual,Soft differential cross-section:

 \begin{eqnarray}
 \frac{d\sigma^{VS}}{dc} & = & C_{V}^{i}\left\{\frac{d\sigma^{WC}}{dc}
 +\frac{d\sigma^{EC}}{dc}\left[\exp(\beta_{e}\ln\frac{1}{y_{0}})-1
 \right]\right\} \nonumber
 \\ &   & \mbox{} +\frac{d\sigma^{EC}}{dc}\left[
 C_{V}^{f}+(\beta_{int}+\beta_{f})\ln\frac{1}{y_{0}}\right]+
 \sum_{i=1}^{N}\frac{d\sigma_{i}^{EC}}{dc}\delta_{i}^{EBOX}
\end{eqnarray}

where

 \begin{eqnarray*}
 \beta_{e} & = & \frac{2\alpha}{\pi}\left[\ln\frac{s}{m_{e}^{2}}-1
 \right]\\
 \beta_{f} & = & \frac{2\alpha}{\pi}\left[\ln\frac{s}{m_{f}^{2}}-1
 \right]\\
 \beta_{int} & = & \frac{4\alpha}{\pi}\ln\frac{t}{u}\\
 C_{V}^{i} & = & 1+\frac{\alpha}{\pi}\left[\frac{3}{2}\ln(\frac{s}
{m_{e}^{2}})+\frac{\pi^{2}}{3}-2\right]+\frac{9}{8}(\frac{\alpha}{\pi}
)^{2}\ln^{2}(\frac{s}{m_{e}^{2}})-\frac{\pi^{2}}{12}\beta_{e}^{2}
 \\
 C_{V}^{f} & = & 1+\frac{\alpha}{\pi}\left[\frac{3}{2}\ln(\frac{s}
{m_{f}^{2}})+\frac{\pi^{2}}{3}-2\right]\\
 m_{f} & = & \mbox{mass of final state leptons} \end{eqnarray*}

The differential cross section d$\sigma^{EC}$/dc is corrected for the
`running' of $\alpha$ in the s and t-channel photon exchange diagrams by
the replacement:
\begin{equation}
\alpha \rightarrow \alpha (u) = \frac{\alpha}{1+\Pi^{\gamma} (u)}
\end{equation}
where $\alpha$ is the on-shell QED fine structure constant
($\alpha^{-1}$ = 137.036...) and $\Pi^{\gamma}(u)=-\Delta \alpha$ is the photon
proper self energy function, parameterised according to Ref.[27].
The most important virtual weak corrections are also included in
d$\sigma^{EC}$/dc via the replacements ~\cite{x9}
\begin{equation}
G_{\mu} \rightarrow \rho_{e}^{DZ} G_{\mu}, \mbox{\hspace{1cm}}  s_{W}^{2}
\rightarrow \kappa_{e}^{DZ}(s) s_{W}^{2}
\end{equation}
where $G_{\mu}$ is the Fermi constant, determined from the muon
lifetime, and $s_{W} = \sin \theta_{W}$ and $\theta_{W}$ is the weak
mixing angle in the on-shell scheme:
\begin{equation}
s_{W}^{2} \equiv 1 - M_{W}^{2}/M_{Z}^{2} \end{equation}
The replacements (3.3) modify the tree level axial vector and vector
coupling constants:
\begin{eqnarray}
g_{A} & = & \frac{1}{4} \left[\frac{\sqrt{2}G_{\mu}M_{Z}^{2}}{\pi \alpha}
\right]\\
g_{V} & = & g_{A}\left[1-4s_{W}^{2}\right] \end{eqnarray}
Since however d$\sigma^{EC}$/dc is multiplied, in Eqn.(3.1) by the
electromagnetic correction of $O(\alpha)$, the effect of the weak
corrections in these terms is quite negligible ( $ \ll 0.1\% $ ) and
so d$\sigma^{EC}$/dc may be considered to be simply
\underline{E}electromagnetically \underline{C}orrected. The label i
in the last term in Eqn.(3.1) denotes the different partial
cross-sections and interference terms (see for example Refs.[3,10,30])
resulting from the two (four) Feynman diagrams that contribute for
$ f \neq e $ ( $ f = e $ ). $\delta_{i}^{EBOX}$ are the corrections
due to $\gamma \gamma $ and $\gamma$ Z box diagrams ~\cite{x3}.

In (3.1) initial state radiation is exponentiated whereas final state
radiation and initial/final interference effects are calculated only to
O($\alpha$). This simplification is justified by the following argument.
The large logarithmic terms at O($\alpha^{2}$), O($\alpha^{3}$),...
associated with initial state radiation do not cancel in the
cross-section when the VS and `hard' contributions, separated by the
arbitrary cut $y_{0}$ on the scaled photon energy, are added, and give
a sizeable (several $\%$) correction. On the other hand, for final state
radiation, because of the KLN theorem ~\cite{x31} the leading logarithms
cancel exactly in the fully integrated cross-section, and cancel almost
completely when only loose cuts are applied. For the level of accuracy
aimed for in BHAGENE3 ( $\approx 0.5 \%$  fractional error in the cross-
section) it is then sufficient to consider, in cross-section calculations
(both $VS$ and `hard'), final state radiation and initial/final
interference effects \footnote{ See Ref.[32] for a discussion of the
effect of initial/final interference in the
charge asymmetry for fermion pair production ($f \neq e$).} effects to
 O($\alpha$).

The differential cross-section (3.1) is integrated over the angular range
$ c_{MIN} < c < c_{MAX} $ :
\begin{equation}
\sigma^{VS} = \int_{c_{MIN}}^{c_{MAX}}\frac{d\sigma^{VS}}{dc}dc
\end{equation}
The probability that a $VS$ type event is generated is proportional to
$\sigma^{VS}$. The lepton scattering angle $\theta_{l}$ is generated by
first calculating an integrated probability distribution in the form of
a histogram with content $P_{i}$ in bin $i$ where:
\begin{equation}
 P_{i} =\left[ \int_{c_{MIN}}^{c_{i}}\frac{d\sigma^{VS}}{dc}dc\right]/
 \sigma^{VS} \end{equation}
Inverting this distribution by linear interpolation yields a LUT for
$c$ with bin index $j$ :
\begin{equation} c_{j} = f(P_{j}) \end{equation}
$c$ is then generated using:
\begin{equation} c_{j} = f(Rn) \end{equation}
In Eqn(3.10) and subsequently $Rn$ is a random number uniformly
distributed
 in the interval $ 0 \leq Rn \leq 1 $. $c$ is calculated by linear
interpolation in the LUT (3.9) using the bins $k, k+1$, adjacent to
$P = Rn$ ($ P_{k} \leq Rn \leq P_{k+1}$).

A similar technique (see also Appendix
C) is used throughout the program for the generation of one or two
dimensional distributions. The advantages are:

(i) Complete generality.

(ii) 100\% efficiency. Multiple trials, as in the weight rejection
technique are unnecessary.

(iii) Fast generation. The simple operations needed to perform
linear interpolation in a LUT use, in general, much less computing
time than the calculation of a weight.

The main disadvantage is a certain overhead in computing time during the
initialisation phase, when the LUT's are created. This becomes
progressively less important as large samples of events are generated.
The number of bins used in the LUT depends on the steepness of the
function to be generated and the desired accuracy. Because of the sharp
peaking of the angular distribution near $c$=1 in Bhabha scattering, due
to $t$-channel photon exchange, a finer binning of the integrated
probability distribution (3.8) is used in this region. 200 bins are
assigned to each of the angular regions:
\[ c_{MIN} < c <  0.8c_{MAX} \]  and
\[ 0.8c_{MAX} < c <  c_{MAX} \]
In the LUT (3.9), 500 uniformly spaced bins are used.

\SECTION{\bf{Hard Photon Corrections}}

\SUBSECTION{
{Generation of Hard Photon Events}}

The `exact' hard photon cross-section is calculated by first generating
events according to simple factorised differential cross-sections,
valid for collinear initial state or final state radiation, and then
reweighting each event according to the procedure described, for example,
in the first of Refs.[4]. `Exact' is written in quotes because the hard
cross section formula used is not the result of an exact, fixed order,
matrix element calculation, but rather an improved soft approximation
comparable to that of Ref.[33]. From now on this distinction will be
understood and the term `exact' used without quotation marks. The exact
 cross-section formula (see Appendix B) is derived by exponentiating
the initial state radiation terms of the O($\alpha$) hard cross-section
{}~\cite{x4} in such a way that consistency with Eqn.(3.1) is obtained in
the soft photon limit:
\begin{equation}  y = 2 \sum_{i} E_{i}^{\gamma} / \sqrt{s} \ll 1
\end{equation}
The use of such a `soft' approximation for the generation of `hard'
photons, may be justified, in the region of the Z peak, by the small
width of the Z relative to its mass :  $\Gamma_{Z}/M_{Z} \approx 0.03$
which strongly damps out hard initial state radiation. Typically $y_{0}$
in Eqn.(3.1) is chosen to be 0.005, corresponding, at the Z peak, to
$ \sum E_{i}^{\gamma} = 225$  $MeV$, so that the majority of `hard'
photons are indeed soft at the scale of $ \Gamma_{Z} = 2.7$ $GeV$.
Exponentiation is not applied to the final state radiation terms, since
the KLN Theorem~\cite{x31} guarantees the smallness of O($\alpha^{2}$),
O($\alpha^{3}$),... corrections in this case, as already discussed in
Section 2 above. Similarly, in accordance with Refs.[11-13,32] the
initial/final interference is also treated only to O($\alpha$).

The approximate differential cross-sections $d \sigma_{A}^{I}$ ,
$d \sigma_{A}^{F}$  for initial, final state radiation respectively,
are:
\begin{equation}
d \sigma_{A}^{I}  = \frac{\alpha sy}{16 \pi^{3} \kappa_{+} \kappa_{-}}
\frac{d \sigma_{0} (s',t^{\star})}{d(-t^{\star})}
\left[(1+\delta_{V}^{i}) y^{\beta_{e}}-y + \frac{y^{2}}{2}\right]
d \Omega_{+} d \Omega_{\gamma} dy
\end{equation}
\begin{equation}
d \sigma_{A}^{F}  = \frac{\alpha sy}{16 \pi^{3} \kappa'_{+} \kappa'_{-}}
\frac{d \sigma_{0} (s,t)}{d(-t)}
d \Omega_{+} d \Omega_{\gamma} dy
\end{equation}
The notation closely follows that of Ref.[3,4].
4-vectors are defined according to:
\[ e^{+}(p_{+})e^{-}(p_{-})  \rightarrow l^{+}(q_{+})l^{-}(q_{-})
\gamma(k) \]
and then:
\begin{eqnarray}
s = (p_{+}+p_{-})^{2} \mbox{\hspace{1.0cm}} t = (p_{+}-q_{+})^{2}
      \mbox{\hspace{1.0cm}}    & & u = (p_{+}-q_{-})^{2} \nonumber \\
s' = (q_{+}+q_{-})^{2} \mbox{\hspace{1.0cm}} t' = (p_{-}-q_{-})^{2}
      \mbox{\hspace{1.0cm}}    & & u' = (p_{-}-q_{+})^{2} \\
\kappa_{\pm} = p_{\pm} \cdot k \mbox{\hspace{1.5cm}}
\kappa'_{\pm} = q_{\pm} \cdot k \mbox{\hspace{1.5cm}}    & &  \nonumber
\end{eqnarray}
\begin{eqnarray}
d\Omega_{+} = d(\cos \theta_{+}) d \phi_{+} \mbox{\hspace{1.0cm}}
d\Omega_{\gamma} = d(\cos \theta_{\gamma})d \phi_{\gamma}
\mbox{\hspace{1.0cm}}  &   &
\end{eqnarray}

$\theta_{+}$, $\phi_{+}$ are the polar and azimuthal angles of the
$\l^{+}$ relative to the incoming $e^{+}$ direction, while
$\theta_{\gamma}$, $\phi_{\gamma}$ are the polar and azimuthal
angles of the photon relative to the incoming $e^{+}$ ( $\l^{+}$)
directions for initial (final) state radiation. Also in Eqns.(4.2,4.3):
\begin{equation} t^{\star} = \frac{-s'}{2}(1-\cos \theta^{\star})
\end{equation}
\begin{equation}
\delta_{V}^{i}  =  \frac{\alpha}{\pi}\left[\frac{3}{2}\ln(\frac{s}
{m_{e}^{2}})+\frac{\pi^{2}}{3}-2\right] \end{equation}
where $\theta^{\star}$ is the $l^{+}$ scattering angle relative to the
incoming $e^{+}$ direction in the Outgoing Dilepton Rest Frame (ODLR).
$d \sigma_{0}$ is the Born level (uncorrected) differential cross-
section for $ e^{+}e^{-} \rightarrow l^{+}l^{-} $. The probabilities
to generate `initial state' (I) or `final state' (F) events, according
to the approximate models are proportional to the cross-sections
$\sigma_{A}^{I}$, $\sigma_{A}^{F}$ given by integrating Eqns.(4.2,4.3)
respectively, over the angles and energies of the outgoing leptons and
photon:
\begin{equation} \sigma_{A}^{I} = \frac{\alpha}{4 \pi} \ln\left(\frac
{s}{m_{e}^{2}}\right) \int_{y_{MIN}}^{y_{MAX}}\frac{dy}{y} \int_{c_{MIN}
^{\star}}^{c_{MAX}^{\star}} \frac{d \sigma_{0}( s',t^{\star})}{d(-t^{
\star})} \left[ (1+\delta_{V}^{i})y^{\beta_{e}}-y+\frac{y^{2}}{2} \right]
d c^{\star} \end{equation}
where
\begin{eqnarray}
s' = s(1-y) \mbox{\hspace{1.0cm}} t^{\star} = -\frac{s}{2}(1-y)
(1-c^{\star})\mbox{\hspace{1.0cm}}    & & c^{\star} = \cos \theta^{\star}
\nonumber
\end{eqnarray}
\begin{equation} \sigma_{A}^{F} = \frac{\alpha}{4 \pi} \ln\left(\frac
{s}{m_{f}^{2}}\right) \ln\left(\frac{y_{MAX}}{y_{MIN}}\right)
\int_{c_{MIN}^{+}}^{c_{MAX}^{+}} \frac{d \sigma_{0}(s,t)}{d(-t)}
d c^{+} \end{equation}
where
\[ c^{+} = \cos \theta_{+} \]

The angular integrals over the lepton scattering angle may be performed
analytically~\cite{x4,x34}. The results are given below in Appendix A.
The expression for the exact differential cross-section
$d \sigma^{EXACT}$ is very lengthy and may be found in Appendix B.

The approximate cross-section $d \sigma_{A}^{I}$, and the exact cross-
section $d \sigma^{EXACT}$ contain, because they are exponentiated,
contributions from 2,3,.. hard photons, even though they are functions
of the angular variables of a single `photon'. In fact the angular
variables of all, except one, of the photons are already integrated
out. On performing the angular integration for this `photon'
\footnote{Note that the energy of this `photon' is in fact the summed
energy of all real photons.} a distribution differential in only the
total photon energy (the derivative of Eqn.(3.1) with respect to
$y_{0}$, with the replacement $y_{0} \rightarrow y$) is obtained.
The ansatz employed is to use the same average weight for all pure
initial state or all pure final state topologies, and, in the case
of one initial and one final state photon in the same event, to
use the harmonic mean of the initial and final state weights.

The calculation of the average weights $\overline{W}'_{I}$,
[$\overline{W}'_{f}$] for events generated according to Eqn.(4.2),
[(4.3)] proceeds as follows. At the end of the initialisation phase
of the program $N(1,0)$,and $N(0,1)$ events are generated with relative
probabilities proportional to $\sigma_{A}^{I}$ and $\sigma_{A}^{F}$
according to Eqns.(4.2) and (4.3). The details of the event generation
algorithm used in each case are given below in Sections 4.3.1 and 4.4.1.
The average weights are then given by the expressions:
\begin{equation}
\overline{W}'_{I} = \frac{1}{N(1,0)} \sum_{i=1}^{N(1,0)}
\frac{d \sigma^{EXACT}(\alpha_{k}^{i})}{d \sigma_{A}^{I}(\alpha_{k}^{i})
+ d \sigma_{A}^{F}(\alpha_k^i)} \end{equation}
\begin{equation}
\overline{W}'_{F} = \frac{1}{N(0,1)} \sum_{j=1}^{N(0,1)}
\frac{d \sigma^{EXACT}(\alpha_{k}^{j})}{d \sigma_{A}^{I}(\alpha_{k}^{j})
+ d \sigma_{A}^{F}(\alpha_k^j)} \end{equation}
where $\alpha_{k}^{i}$, ($\alpha_{k}^{j}$) are the $k$ kinematical
variables necessary to completely specify the kinematical configuration
of the $i$th I event ($j$th F event). In the subsequent event generation
phase weights are assigned to multiphoton events according to Table 1.
The relative probabilities of radiating 1,2,3 photons are given by a
Poisson distribution function~\cite{x35}:
\begin{eqnarray}
P_{I}^{(\geq 2 \gamma)}/ P_{I}^{(\geq 1 \gamma)}  & = & 1-e^{-r_{e}} \\
P_{F}^{(\geq 2 \gamma)}/ P_{F}^{(\geq 1 \gamma)}  & = & 1-e^{-r_{f}} \\
P_{F}^{(\geq 3 \gamma)}/ P_{F}^{(\geq 1 \gamma)}  & = & 1-(1+r_{f})
e^{-r_{f}} \end{eqnarray}
where
\begin{equation}
r_{e} = \beta_{e} \ln \left( \frac{1}{y_{0}} \right), \mbox{\hspace{1cm}}
r_{f} = \beta_{f} \ln \left( \frac{1}{y_{0}} \right)
\end{equation}
These probabilities are used to construct the `a priori' event generation
probabilities $P(n_{\gamma}^{I},n_{\gamma}^{F})$ presented in Table 2.
The following definitions are used in this Table:
\begin{eqnarray}
P_{VS}  & = & \sigma^{VS}/ \sigma^{A}_{TOT} \\
P_{I}  & = & \sigma_{A}^{I}/ \sigma^{A}_{TOT} \\
P_{F}  & = & \sigma_{A}^{F}/ \sigma^{A}_{TOT}  \end{eqnarray}
where \[ \sigma_{TOT}^{A} = \sigma^{VS} + \sigma_{A}^{I} + \sigma_{A}^
{F} \]
and
\begin{equation}
 \rho_{IF} = (S_{I}+S_{F})/(S_{I}+S_{F}+\sqrt{S_{I}S_{F}})\end{equation}
where
\begin{equation}
S_{I} = P_{I}(1-e^{-r_{e}}), \mbox{\hspace{1cm}}
S_{F} = P_{F}(1-e^{-r_{f}})
\end{equation}
In Table 2 $P(2,0)$, $P(0,3)$ are actually the probabilities for
$ \geq 2$, ($ \geq 3$) initial, (final) state photons, so so that initial
states with 2,3,.. photons and final states with 3,4,.. photons are
assigned exactly 2 and 3 photons respectively. $P(1,1)$ is derived from
$P(2,0)$ and $P(0,2)$ using a factorisation ansatz. The probabilities
in Table 2 sum to unity.
For the case of single photon I or F events the weight $W$ is calculated
as:
\begin{equation}
W =
\frac{d \sigma^{EXACT}(\alpha_{k})}{d \sigma_{A}^{I}(\alpha_{k})
+ d \sigma_{A}^{F}(\alpha_k)} \end{equation}
The generation of the kinematical variables $\alpha_{k}$ necessary to
completely define the event configuration is described below.
After checking for consistency with the imposed kinematical cuts
(events failing the cuts are assigned weight zero) events with unit
weight are written out by requiring that:
\begin{equation} W/W_{MAX} > Rn  \end{equation}
$W_{MAX}$ is chosen as small as possible (typically $W_{MAX} = 2$) in
order to maximise the efficiency of event generation.

\SUBSECTION{
{Cross-Section Calculation}}

The calculation of the cross-section printed out after completion of
the event generation loop is now described. Suppose that $N$ event
generation trials are made in the loop with $N(n_{\gamma}^{I},
n_{\gamma}^{F})$ trials in each different final state channel
chosen according to the probabilities in Table 2.
Event weights are defined as follows :
\begin{eqnarray}
W_I^i & = &
\frac{d \sigma^{EXACT}(\alpha_{k}^i)}{d \sigma_{A}^{I}(\alpha_{k}^i)
+ d \sigma_{A}^{F}(\alpha_k^i)}, \mbox{\hspace{1cm}} 1 < i < N(1,0) \\
W_F^j & = &
\frac{d \sigma^{EXACT}(\alpha_{k}^j)}{d \sigma_{A}^{I}(\alpha_{k}^j)
+ d \sigma_{A}^{F}(\alpha_k^j)}, \mbox{\hspace{1cm}} 1 < j < N(0,1)  \\
W_{VS}^l & = & 1,
 \mbox{\hspace{4cm}} 1 < l < N(0,0)  \end{eqnarray}
where the $\alpha_k^{i,j}$ are defined after Eqns.(4.10,4.11).
The cross-section $\sigma^{CUT}$ corresponding to the imposed
kinematical cuts (trial events that fail the cuts are assigned weight
zero) is given by:
\begin{equation}
\sigma^{CUT} = \overline{W}(\sigma^{V,S}+\sigma_A^I+\sigma_A^F)
\end{equation}
where

\begin{eqnarray}
 \overline{W} & = & [(N(1,0)+N(2,0)\overline{W}_I+(N(0,2)+N(0,2)+N(0,3))
\overline{W}_F  \nonumber \\
              &   & +N(1,1)\sqrt{\overline{W}_I \overline{W}_F}+
N(0,0) \overline{W}_{VS}]/N
\end{eqnarray}
\begin{eqnarray}
 \overline{W}_I  & = & \frac{1}{N(1,0)} \sum_{i=1}^{N(1,0)} W_I^i \\
 \overline{W}_F  & = & \frac{1}{N(0,1)} \sum_{j=1}^{N(0,1)} W_F^j \\
 \overline{W}_{VS}  & = & \frac{1}{N(0,0)} \sum_{l=1}^{N(0,0)}
W_{VS}^l \end{eqnarray}

The weights $\overline{W}_I$, $\overline{W}_F$ defined by (3.26), (3.27)
are identical to $\overline{W}'_{I}$, $\overline{W}'_f$ defined by
(4.10,4.11). Since however, the former are more precise than the
latter, due to the typically smaller statistical errors of the
Monte-Carlo integration in the main generation loop, they are preferred
for the calculation of $\sigma^{CUT}$. The error on $\sigma^{CUT}$,
$\Delta \sigma^{CUT}$ due to the statistical error of the Monte-Carlo
integration is~\cite{x36}:
\begin{equation}
\frac{\Delta \sigma^{CUT}}{\sigma^{CUT}} = [SW2-(SW)^2/N_{IF}]^{\frac
{1}{2}}/SW \end{equation}
where
\begin{eqnarray*}
SW2  & = & \sum_{i=1}^{N(1,0)}(W_I^i)^2+\sum_{j=1}^{N(0,1)}(W_F^j)^2 \\
SW  & = & N(1,0)\overline{W}_I+N(0,1)\overline{W}_F \\
N_{IF}  & = & N(1,0)+N(0,1)  \end{eqnarray*}

\SUBSECTION{
{Initial State Radiation Events}}

\SUBSUBSECTION{\tt{One Initial State Photon}}

The final state 4-vectors of events with a single initial state photon
are generated by the subroutine ZINIGB. These events are first generated
according to the approximate differential cross-section (4.2) and then
re-weighted according to the exact differential cross-section using
Eqn.(4.23).

During the initialisation phase a LUT for the scaled photon energy
$y = E_{\gamma}/E$, with 200 uniformly spaced bins from $y_{MIN}$ to
$y_{MAX}$ is created, as described in Section 2 above. This is done by
integrating Eqn.(4.2) first over $\cos\theta^\star$, then over $y$. The
angular integral is done analytically (see Appendix A), and the y
integral by numerical integration \footnote{Using the CERN
Library program DGAUSS}. At the same time a two dimensional LUT
(200 bins in $y$, 500 in $\cos\theta^{\star}$) is created. Details of the
creation and use of such a two dimensional LUT may be found in Appendix
C. The subroutines used to create these one dimensional (1-D) and two
dimensional (2-D) LUT are SETYD, SETCTD respectively.

The kinematical variables $\alpha_{k}$ used to completely specify
the configuration are, in order of their generation:

  \[ y,\mbox{\hspace{0.5cm}} \cos\theta_{\gamma},\mbox{\hspace{0.5cm}}
  \phi_+^{\star},
  \mbox{\hspace{0.5cm}}\cos\theta^{\star},\mbox{\hspace{0.5cm}}
  \phi_{\gamma} \]

The scaled photon energy y is chosen from the 1-D LUT. The angle
$\theta_{\gamma}$ between the incoming $e^{+}$ and the photon is chosen
according to the distribution:
\begin{equation}
\frac{dn}{dc_{\gamma}} \simeq \frac{1}{1-\beta_{IN}^{2}c_{\gamma}^{2}}
\end{equation}
where
  \[ c_{\gamma} = \cos \theta_{\gamma},\mbox{\hspace{1.5cm}}
   \beta_{IN} = \sqrt{1-\frac{4m_e^2}{s}} \]
In this case the integrated probability distribution is calculated
analytically and inverted to give:
\begin{equation}
c_{\gamma} = (a-1)/[\beta_{IN}(a+1)] \end{equation}
where
\[ a = [(1-\beta_{IN})/(1+\beta_{IN})]\exp [2Rn\ln \frac{1+\beta_{IN}}
 {1-\beta_{IN}}] \]

The azimuthal angle $\phi_+^{\star}$ between the planes P1 (defined by
photon and the incoming $e^{+}$), and P2 (defined by the outgoing $l^+$
and the incoming $e^+$), in the outgoing dilepton rest frame (ODLR),
is generated uniformly. Next the 2-D LUT is used, together with the
previously generated value of y to give $\cos\theta^{\star}$. Finally
the angle $\phi_{\gamma}$, giving the orientation of the plane defined
by the photon and the incoming $e^+$ in the incoming $e^+,e^-$ (LAB)
frame, is generated uniformly. To obtain the 4-vectors in the LAB
system of the outgoing $l^+,l^-,\gamma$ a Lorentz transformation
between the ODLR and the LAB frames is required. The corresponding
boost direction is parallel to the photon direction. Denoting by
$\alpha_B$ the angle between the photon and the incoming $e^+$ in
the ODLR frame, the transformation gives:

\begin{eqnarray}
\sin \alpha_B & = & \frac{ \sin \theta_{\gamma}}{1-\beta^{\star}
\cos\theta_{\gamma}} \\
\cos \alpha_B & = & \frac{ \cos \theta_{\gamma}-\beta^{\star}}
{1-\beta^{\star}
\cos\theta_{\gamma}} \end{eqnarray}

where
\begin{equation}
 \beta^{\star} = y/(2-y) \end{equation}

The 3-momentum components of the $l^+$ in the ODLR frame:
\begin{eqnarray}
 q_+^{(1)\star} & = & q^{\star}s^{\star} \sin \phi_+^{\star} \\
 q_+^{(2)\star} & = & q^{\star}s^{\star} \cos \phi_+^{\star} \\
 q_+^{(3)\star} & = & q^{\star}c^{\star} \end{eqnarray}
where
  \[ s^{\star}  = \sin \theta^{\star},\mbox{\hspace{1.0cm}}
   c^{\star} = \cos \theta^{\star},\mbox{\hspace{1.0cm}}
   q^{\star} = \sqrt{\frac{s(1-y)}{4}-m_l^2} \]
 are transformed into the LAB frame using the Lorentz transformation
defined by Eqns.(4.34-4.36). Conservation of 3-momentum in the LAB
frame (the 3-momenta of the $l^+$ and photon being known) then determines
the 3-momentum of the $l^-$. Energy conservation of the complete event is
now checked. In the case of a deviation of more than $\pm 0.5\%$ the
event is rejected and a new one is generated. A similar check is made
for all other hard photon event topologies described in the following
Sections. If the event is accepted, the 3-vectors of all particles
are rotated about the incoming $e^+$ axis, so that the azimuthal
angle of the photon becomes $\phi_{\gamma}$.

\SUBSUBSECTION{\tt{Two Initial State Photons}}

Events with two initial state photons are generated by subroutine
ZINI2G according to an improved soft photon approximation
{}~\cite{x3,x14,x37}. The differential cross-section is:
 \begin{eqnarray}
 d\sigma^I_{2\gamma} & = & C_{n} d\sigma_{0}(s')
 \frac{p_{+} \cdot p_{-}}{(p_{+} \cdot k_{1})
 (p_{-} \cdot k_{1})}\frac{p_{+} \cdot p_{-}}{(p_{+} \cdot k_{2})(
 p_{-} \cdot k_{2})}\nonumber \\
   &   & \mbox{} \times (1-\frac{k_{1}}{E}+\frac{k_{1}^{2}}{2E^{2}})
(1-\frac{k_{2}}{E}+\frac{k_{2}^{2}}{2E^{2}}) \frac{d^{3}k_{1}}{k_{1}}
\frac{d^{3}k_{2}}{k_{2}} \nonumber \\
  &   & \nonumber \\
  & = & 4C_{n}\frac{d\sigma_{0}(s') d(y_{1}) d(y_{2}) dy_{1}
d\Omega_{1} dy_{2}d\Omega_{2}}{(1-\beta_{IN}^{2}\cos^{2}\Theta_
{\gamma 1})(1-\beta_{IN}^{2}\cos^{2}\Theta_{\gamma 2})}
\end{eqnarray}
where
\[
s' = M_{l^+l^-}^2,\mbox{\hspace{0.5cm}} E =\sqrt{s}/2,\mbox{\hspace{0.5cm}}
y_i = 2 k_i/\sqrt{s}, \mbox{\hspace{0.5cm}}  d(y)=(1-y+
\frac{y^{2}}{2})/y
\]

Here $d\sigma_0(s')$ is the Born differential cross-section and $k_1$,
$k_2$;$d\Omega_1$,$d\Omega_2$ are, respectively, the energies and solid
angle elements of the radiated photons. $\theta_{\gamma 1}$,
$\theta_{\gamma 2}$ are the angles between the photons and the incoming
$e^+$. The factors $d(y_i)$ are the Gribov-Lipatov~\cite{x38} kernels.
In Eqn.(4.40) recoil effects are neglected, whereas in the event
generation algorithm,  the change in direction of the radiator due to
the recoil from the first radiated photon (in the case that both photons
are radiated from the same incoming $e^{\pm}$) is allowed for. The photon
angles relative to the radiator are still however generated according to
Eqn.(4.40), so that effects due to the virtuality of the first
recoiling $e^{\pm}$ are not taken into account. Also symmetrisation
effects due to different possible time ordering in the radiation of
two photons from the same incoming line are neglected.

The kinematical variables $\alpha_k$ used to define the configuration of
an event with two initial state photons are, in order of generation:

  \[ y,\mbox{\hspace{0.5cm}} \cos\theta_{\gamma 1},\mbox{\hspace{0.5cm}}
  \cos\theta_{\gamma 2},\mbox{\hspace{0.5cm}} u,
  \mbox{\hspace{0.5cm}} \phi_{\gamma 12},\mbox{\hspace{0.5cm}}
  \phi_+^{\star}, \mbox{\hspace{0.5cm}} \cos \theta^{\star},
  \mbox{\hspace{0.5cm}} \phi_{\gamma 1} \]

Here $y$ is the scaled total photon energy:
\begin{equation}
 2y_{MIN} \leq y = 2(k_1+k_2)/\sqrt{s} \leq 2y_{MAX}
\end{equation}
$u$ is defined as $\ln(y_1/y_2)$ where $y_1$ and $y_2$ are the scaled
 energies of the photons:
\begin{equation}
 y_{MIN} \leq y_i = 2k_i/\sqrt{s} \leq y_{MAX}
\end{equation}
and $y_1 > y_2$. The angles $\theta_{\gamma 1}$, $\theta_{\gamma 2}$
are generated according to the same distribution (Eqn.(4.32)) as the
angle $\theta_{\gamma}$ for the case of single initial state photon
events.
 $\phi_{\gamma 12}$ is the angle between the planes
defined by the incoming beam direction and each of the photons.
The angle $\phi_+^{\star}$ is that between the plane formed by the
summed momentum vector of the two photons and the beam direction,
and the plane formed by the $l^+$ and the beam direction, all in the ODLR
frame. The other variables are defined in the same way as for single
initial state radiation events.
The variables $y$ and $u$ are chosen using LUT generated at
initialisation.
The joint distribution in $y$ and $u$ is derived from that for $y_1$ and
$y_2$ (Eqn.(4.40)):
\begin{equation}
\frac{d^2n}{dy_1dy_2} \simeq \int_{c_{MIN}^{\star}}^{c_{MAX}^{\star}}
d\sigma_0(\tilde{s}',c^{\star})dc^{\star}d(y_1)d(y_2)
\end{equation}
where:
\begin{equation}\tilde{s}' = s(1-y_1-y_2+y_1y_2) \simeq s'\end{equation}
In general the outgoing dilepton effective mass $\sqrt{s'}$ depends also
on the angles of the radiated photons \footnote{The exact expression, in
the present case, is given after Eqn.(4.57) below}.
 The expression (4.44) holds for
back-to-back photons, the configuration corresponding to the maximum
value of $y$ in (4.40), and therefore spanning the full possible phase
space. For collinear photons, where $y_{MAX} = 1$, the $y_1y_2$ term in
(4.44) is absent. Separate LUT are thus generated for events with
(approximately) back-to-back or collinear photons. If $\cos \theta_{\gamma 1}$,
  $\cos \theta_{\gamma 2}$ have opposite signs (back-to-back configuration)
  then Eqn.4.44 is used to define $\tilde{s}'$ in Eqn.4.44 and $y_{MAX}
  \simeq 2$. If  $\cos \theta_{\gamma 1}$ , $\cos \theta_{\gamma 2}$
  have the same sign (collinear configuration), the $y_1y_2$ term in
  Eqn 4.44 is omitted and $y_{MAX} \simeq 1$.

By the change of variables:
\begin{eqnarray}
 y_1  & = & \frac{y}{2}(1+R) \\
 y_2  & = & \frac{y}{2}(1-R) \\
 R    & = & (e^u-1)/(e^u+1)  \\
 W    & = & \ln(1/y) \end{eqnarray}
 (4.43) may be re-written as:
\begin{equation}
\frac{d^2n}{dWdu} \simeq \int_{c_{MIN}^{\star}}^{c_{MAX}^{\star}}
d\sigma_0(\tilde{s}',c^{\star})dc^{\star}d_1d_2
\end{equation}
where:

  \[ d_i = 1-y_i+\frac{y_i^2}{2},\mbox{\hspace{0.5cm}}
   y_1 = e^{u-W}/(e^u+1),
  \mbox{\hspace{0.5cm}} y_2 = e^{-W}/(e^u+1) \]

 1-D LUTs for $y=e^{-W}$ are created at initialisation by subroutines
SETYDJ,SETYDL for back-to-back, collinear photons respectively
 using Eqn.(4.49). The integral over $c^{\star}$ is performed
analytically, and those over $W$ and $u$ numerically, using, respectively
Simpson's rule and Gaussian integration. At the same time 2-D LUTs in
the variables $W$ and $u$ are created by the subroutines SETRI, SETRL
for back-to-back, collinear photons respectively.
The integration
limits on $W = \ln(1/y)$ are derived from Eqn.(4.41).
Overall energy
conservation leads to the following limits for the variable $u$ :
\begin{eqnarray}
u_{MAX} & = & \ln [(y+y_1^{MAX}-y_2^{MIN})/(y-y_1^{MAX}+y_2^{MIN})]
\\
u_{MIN} & = & -u_{MAX} \end{eqnarray}
where:
\begin{eqnarray*}
y_1^{MAX} & = & \min [ y_{MAX},y-y_{MIN} ] \\
y_2^{MIN} & = & \max [ y_{MIN},y-y_{MAX} ]  \end{eqnarray*}

Because the factorised formula (4.49) used to generate the
photon energies does not take into account the angles of the radiated
photons, overall overall energy conservation of the event is checked,
after generation of the variables $\alpha_k$ and before construction
of 4-vectors. In the LAB system, the 3-momentum of the outgoing
dilepton is equal to the total 3-momentum of the two photons. Energy
conservation then requires that:
\begin{eqnarray}
 \sqrt{s} & = & k_1+k_2 +[(\vec{k_1}+\vec{k_2})^2+M_{l^+l^-}^2]^\frac
 {1}{2} \\
          & > & k_1+k_2 +|\vec{k_1}+\vec{k_2}| \end{eqnarray}
 The inequality (4.53) follows from the condition that the dilepton
effective mass $M_{l^+l^-}$ is $ > 0 $. (4.53) may also be written as:
\begin{eqnarray}
 2 > &  &y_1 + y_2  +  [ y_1^2 +y_2^2
 + 2y_1y_2(\sin \theta_{\gamma 1}\sin \theta_{\gamma 2}
 \sin \phi_{\gamma 12}+\cos \theta_{\gamma 1} \theta_{\gamma 2}]^\frac{1}
 {2}  \end{eqnarray}
 If the R.H.S. of (4.54) is found to be $ > 1.99 $ the current event is
 rejected and a new one generated.

 In constructing the 4-vectors of the photons, the recoil of the
 radiating $e^{\pm}$ is allowed for. The recoil angle, $\alpha_R$ of
 the $e^{\pm}$, after photon 1 has been radiated, is:
\begin{equation}
\alpha_R = \arctan \left( \frac{y_1 \sin \theta_{\gamma 1}}
{1-y_1| \cos \theta_{\gamma 1}|} \right) \end{equation}
In the case of radiation of the two photons from the same incoming
line, given by the condition:

 \[ \cos \theta_{\gamma 1} \cos \theta_{\gamma 2} > 0 \]

the angle $\theta_{\gamma 2}$ is chosen relative to the direction
of the recoiling $e^{\pm}$, rather than the beam direction.

The remainder of the event generation follows closely the procedure
for the case of one initial state photon, described above. The lepton
scattering angle is chosen using the same 2-D LUT as in the one photon
case. The boost between th LAB and the ODLR frames is now along the
direction of the vector sum ($ \vec{K}$) of the 3-momenta of the two
photons. The angle $\theta_{\gamma}$ in Eqns.(4.34,4.35) is replaced
by $\theta_{K}$, the angle between the incoming $e^+$ direction and
$\vec{K}$. The Lorentz transformation parameter $\beta^{\star}$ of
Eqn.(4.36) is replaced by the expression:
\begin{equation}
\beta^{\star} =  \frac{2| \vec{K} |}{\sqrt{s}(2-y)}
\end{equation}
 and the lepton momentum in the ODLR frame used in Eqns.(4.37-4.39)
 is now given by:
\begin{equation}
 q^{\star} = \sqrt{\frac{s'}{4}-m_l^2}
\end{equation}
where
 \[ s' = \frac{s}{4}(2-y)^2-(\vec{K})^2   \]
 Finally, since events were generated using the approximate value of $s'$,
 $\tilde{s}'$ given by Eqns.(4.43) and (4.44) (with the appropriate
 modification for collinear photons) the events are re-weighted using a
 WRP, to take into account the exact value of $s'$ as given above
 after Eqn.(4.57). The weight used is:
 \[ d \sigma_{0}(s',c^{*}) / d \sigma_{0}(\tilde{s}',c^{*}) \]

\SUBSECTION{
{Final State Radiation Events}}

\SUBSUBSECTION{\tt{One Final State Photon}}

Events with a single final state photon are generated by the subroutine
ZFINGB, according to the differential cross-section Eqn.(4.3). They
are subsequently re-weighted according to the exact differential
cross-section using Eqn.(4.24)

The kinematical variables $\alpha_{k}$ used to define the event
configuration are, in order of their generation:

  \[ y,\mbox{\hspace{0.5cm}} \cos\theta_{\gamma}^{\star},
  \mbox{\hspace{0.5cm}} \cos \theta_{\pm},
  \mbox{\hspace{0.5cm}}
  \phi_{\gamma}',
  \mbox{\hspace{0.5cm}}
  \phi_{\pm} \]

Following (4.8) the scaled photon energy, $y$, is given by:
\begin{equation}
 y = y_{MIN} \exp[ Rn \ln \left( \frac{y_{MAX}}{y_{MIN}} \right)]
\end{equation}
The angle, $\theta_{\gamma}^{\star}$, between the $l^+$ and the photon in
the ODLR frame, is chosen according to the distribution:
\begin{equation}
 \frac{dn}{dc_{\gamma}^{\star}} \simeq \frac{1}{1-\beta_{FI}^2
(c_{\gamma}^{\star})^2} \end{equation}
where
  \[ c_{\gamma}^{\star} = \cos \theta_{\gamma}^{\star},
  \mbox{\hspace{1.5cm}}
   \beta_{FI} = \sqrt{1-\frac{4m_l^2}{s}} \]

The scaled energies of the $l^+,l^-$ in the LAB frame, $y_+,y_-$, are
obtained by a Lorentz transformation:
\begin{equation}
y_{\pm} = \frac{2}{\sqrt{s}} \gamma^{\star}
[q^{(0)\star} \pm \beta^{\star} q^{\star}
c_{\gamma}^{\star}]
\end{equation}
where
\begin{eqnarray*}
   q^{(0)\star}  = \frac{\sqrt{s(1-y)}}{2},
  \mbox{\hspace{1.5cm}}&  &
   q^{\star}  = \sqrt{(q^{(0)\star})^2-m_l^2} \\
   \beta^{\star} = y/(2-y),
  \mbox{\hspace{1.5cm}}&  &
   \gamma^{\star}  = (1-\frac{y}{2})/\sqrt{1-y}
\end{eqnarray*}

The LAB scattering angles $\theta_{\pm}$ of the $l^{\pm}$ are found
using the same 2-D LUT in $y$ and $\cos \theta^{\star}$ as used for initial
state radiation. Here $y$ is set to zero so that $s' \rightarrow s$
and $\theta^{\star} \rightarrow \theta_+$. Following Ref[14], the angle
$\theta_+$ is assigned to the $l^+$, or $\pi-\theta_+$ to the $l^-$,
according to whether, or not :
\begin{equation} P_s > Rn \end{equation}
where
\[   P_s = y_+^2/(y_+^2+y_-^2) \]
This has the effect that, if the photon is radiated from the $l^{\pm}$
then the scattering angle is most likely assigned to the other lepton
$l^{\mp}$, whose direction is unaffected by recoil effects. In the case
that the $\theta_+$ is assigned to the $l^+$, the 3-axis is chosen along
the $l^+$ direction and the 3-momentum components of the $l^+$,$\gamma$
are given, in the LAB system, by the expressions:
\begin{eqnarray}
 q_+^{(1)} & = & 0 \nonumber \\
 q_+^{(2)} & = & 0  \\
 q_+^{(3)} & = & \frac{\sqrt{s}}{2} y_+ \beta_+ \nonumber \end{eqnarray}
\begin{eqnarray}
 k^{(1)} & = &  \frac{\sqrt{s}}{2} y s_{\gamma}' \cos \phi_{\gamma}'
 \nonumber \\
 k^{(2)} & = &  \frac{\sqrt{s}}{2} y s_{\gamma}' \sin \phi_{\gamma}' \\
 k^{(3)} & = &  \frac{\sqrt{s}}{2} y c_{\gamma}'
 \nonumber \end{eqnarray}
where
  \[ c_{\gamma}'  = 1-2[1-(1-y)/y_+]/y,\mbox{\hspace{1.0cm}}
   s_{\gamma}' = \sqrt{1-(c_{\gamma}')^2},\mbox{\hspace{1.0cm}}
   \beta_+ = \sqrt{1-\frac{4m_l^2}{s y_+^2}} \]

Here $\phi_{\gamma}'$ is the azimuthal angle of the photon about the
$l^+$ direction. If $\pi-\theta_+$ is assigned to the $l^-$, the
3-axis is chosen along the $l^-$ direction, and the 3-momentum
components of the $l^-$, $\gamma$ are given in the LAB system as:
\begin{eqnarray}
 q_-^{(1)} & = & 0 \nonumber \\
 q_-^{(2)} & = & 0  \\
 q_-^{(3)} & = & \frac{\sqrt{s}}{2} y_- \beta_- \nonumber \end{eqnarray}
\begin{eqnarray}
 k^{(1)} & = &  \frac{\sqrt{s}}{2} y s_{\gamma}''\cos \phi_{\gamma}'
 \nonumber \\
 k^{(2)} & = &  \frac{\sqrt{s}}{2} y s_{\gamma}''\sin \phi_{\gamma}' \\
 k^{(3)} & = &  \frac{\sqrt{s}}{2} y c_{\gamma}''
 \nonumber \end{eqnarray}
where
  \[ c_{\gamma}'' = 1-2[1-(1-y)/y_-]/y,\mbox{\hspace{1.0cm}}
   s_{\gamma}'' = \sqrt{1-(c_{\gamma}'')^2},\mbox{\hspace{1.0cm}}
   \beta_- = \sqrt{1-\frac{4m_l^2}{s y_-^2}} \]

The momentum components in the standard LAB system with 3-axis along
the incoming $e^+$ direction, are found by rotating the above 3-vectors
by $\theta_+$ (or $\pi- \theta_+$) about the 1-axis, as appropriate.
The azimuthal angle $\phi_+$,($\phi_-$) of the $l^+$, ($l^-$) about the
3-axis is then generated uniformly. Finally the 3-momentum of the
remaining lepton is given by momentum conservation in the LAB frame.

\SUBSUBSECTION{\tt{Two Final State Photons}}

Events with two final state photons are generated by the subroutine
ZFIN2G according to the factorised differential cross-section
{}~\cite{x3,x14,x37}:
 \begin{eqnarray}
 d\sigma^F_{2\gamma} & = & C_{n} d\sigma_{0}(s)
 \frac{q_{+} \cdot q_{-}}{(q_{+} \cdot k_{1})
 (q_{-} \cdot k_{1})}\frac{q_{+} \cdot q_{-}}{(q_{+} \cdot k_{2})(
 q_{-} \cdot k_{2})}\nonumber \\
   &   & \mbox{} \times (1-\frac{k_{1}}{E}+\frac{k_{1}^{2}}{2E^{2}})
(1-\frac{k_{2}}{E}+\frac{k_{2}^{2}}{2E^{2}}) \frac{d^{3}k_{1}}{k_{1}}
\frac{d^{3}k_{2}}{k_{2}} \nonumber \\
  &   & \nonumber \\
  & = & 4C_{n}\frac{d\sigma_{0}(s) d(y_{1}) d(y_{2}) dy_{1}
d\Omega_{1} dy_{2}d\Omega_{2}}{(1-\beta_{FI}^{2}\cos^{2}\Theta_
{\gamma 1}')(1-\beta_{FI}^{2}\cos^{2}\Theta_{\gamma 2}')}
\end{eqnarray}
The variables in (4.66) are defined after Eqns.(4.40) and (4.59), with
the exception of $\theta_{\gamma 1}'$, $\theta_{\gamma 2}'$, which are
the angles between the photons 1,2 and the $l^+$ in the LAB system. To
take into account recoil effects however, when $\cos \theta_{\gamma 2}'
 < 0$ the photon 2 is assigned the angle $\pi-\theta_{\gamma 2}'$
relative to the $l^-$ direction rather than $\theta_{\gamma 2}'$
relative to the $l^+$ direction.

The kinematical variables $\alpha_{k}$ in the order of generation are:

  \[ y,\mbox{\hspace{0.5cm}} cos\theta_{\gamma 1}',\mbox{\hspace{0.5cm}}
  cos\theta_{\gamma 2}',\mbox{\hspace{0.5cm}} u,
  \mbox{\hspace{0.5cm}}cos\theta_{\pm},\mbox{\hspace{0.5cm}}
  \phi_{\gamma 12}', \mbox{\hspace{0.5cm}} \phi_{\gamma T}',
  \mbox{\hspace{0.5cm}} \phi_{\pm} \]

Here $\theta_{\pm}$ are
the scattering angles of the $l^{\pm}$ relative to the incoming $e^+$
in the LAB frame. The azimuthal angle $\phi_{\gamma12}'$ is chosen to
be that between the planes defined by : $(\vec{k_1}, \vec{q_+})$;
$(\vec{k_2}, \vec{q_+})$ for $\cos \theta_{\gamma 2}' > 0$ and between
the planes: $(\vec{k_1}, \vec{q_-})$; $(\vec{k_2}, \vec{q_-})$ for
$\cos \theta_{\gamma 2}' < 0$. $\phi_{\gamma T}'$ is the azimuthal
angle about the $l^+$ direction of the total momentum vector
$\vec{K_2}$ of the two photons. $\phi_{\pm}$ are the azimuthal angles
of the $l^{\pm}$ relative to the incoming $e^+$. All three angles
are generated uniformly in the interval 0 to $2 \pi$ radians. The
polar angles $\theta_{\gamma 1}'$, $\theta_{\gamma 2}'$ are generated
according to Eqn.(4.32) with the replacements $\theta_{\gamma}
\rightarrow \theta_{\gamma}'$, $\beta_{IN} \rightarrow \beta_{FI}$.
The variables $y$ and $u$ ($ y = y_1+y_2$, $u =\ln (y_1/y_2)$) are chosen
from, respectively, 1-D and 2-D LUT's generated at initialisation.
The distribution (4.49) is here replaced by:
\begin{equation}
\frac{d^2n}{dWdu} \simeq d_1d_2
\end{equation}
where $W = \ln(1/y)$. The subroutines that generate the 1-D LUT for
$y$ and the 2-D LUT for $W$ are SETYDK, SETRK respectively. The angle
$\theta_+$ is generated as described above for single final state
photon events.

The construction of the 3-vectors of the final state particles is
now described. Unlike for the case of single final state photon
radiation it is here more convenient to work entirely in the LAB
frame. Choosing the 3-axis along the $l^+$ direction, with the
2-axis in the plane containing $\vec{q_+}$ and $\vec{k_1}$ the
momentum components of photon 1 are:
\begin{eqnarray}
 k^{(1)}_1 & = &  0
 \nonumber \\
 k^{(2)}_1 & = &  \frac{\sqrt{s}}{2} y_1 s_{\gamma 1}' \\
 k^{(3)}_1 & = &  \frac{\sqrt{s}}{2} y_1 c_{\gamma 1}'
 \nonumber \end{eqnarray}
where

  \[ s_{\gamma 1}'  = \sin \theta_{\gamma 1}',\mbox{\hspace{1.0cm}}
   c_{\gamma 1}'  = \cos \theta_{\gamma 1}' \]

In the case that
$c_{\gamma 2}' > 0$ (photon 2 radiated from the $l^+$) the momentum
components of photon 2 are given by:
\begin{eqnarray}
 k^{(1)}_2& = &\frac{\sqrt{s}}{2}y_2 s_{\gamma 2}'\cos \phi_{\gamma 12}'
 \nonumber \\
 k^{(2)}_2& = &\frac{\sqrt{s}}{2}y_2 s_{\gamma 2}'\sin \phi_{\gamma 12}'
  \\
 k^{(3)}_2& = &\frac{\sqrt{s}}{2}y_2 c_{\gamma 2}'
 \nonumber \end{eqnarray}
For $c_{\gamma 2}' < 0$ (photon 2 radiated from the $l^-$) the direction
of the 3-axis in Eqn.(4.69) is chosen opposite to the direction of the
$l^-$. In this case the momentum components are still constructed
according to Eqn.(4.69), but they are subsequently rotated about the
1-axis through the acollinearity angle $\alpha_{A}$ between the $l^+$,
 $l^-$ directions. This angle is calculated from the equations:
\begin{eqnarray}
 \sin \alpha_A & = & y_1 s_{\gamma 1}' / y_- \nonumber \\
 y_-  & = & 2-y_1-(1-y_1)/[1-0.5(1-c_{\gamma 1}') y_1] \\
 \cos \alpha_A & = & \sqrt{1-\sin^2 \alpha_A}  \nonumber
\end{eqnarray}
By this procedure the strong peaking of the radiated photons along the
$l^{\pm}$ directions is properly accounted for, even in the case of
recoils generated by the first radiated photon \footnote{ In Eqn.(4.69)
the angle $\theta_{\gamma 2}'$ is that between the photon and the
radiating (virtual) $l^-$, rather than that between the photon and the
outgoing $l^-$ as in Eqn.(4.66). This difference is of little importance
for soft or nearly collinear photons. For hard non-collinear photons
however the rate will be over-estimated as compared to Eqn.(4.66). In
the case that both photons are radiated from the $l^+$ the ansatz used
tends, on the contrary to underestimate the rate of such photons as
compared to Eqn.(4.66)}. Before proceeding to construct the 3-vectors
of the outgoing leptons energy conservation is checked using Eqn.(3.51).
If the R.H.S. is $ > 0.99 \sqrt{s}$ the event is rejected and a new one
is generated.

The scaled energies $y_{\pm}$ of the $l^{\pm}$ are next calculated:
\begin{eqnarray}
y_+ & = & 2[1-y(1-\frac{y}{4})-\frac{(\vec{K_2})^2}{s}]
/(2-y+\frac{2K_2^{(3)}}{\sqrt{s}}) \\
y_- & = & 2-y-y_+ \end{eqnarray}
As in Eqns.(4.68,4.69) the 3-axis is along the $l^+$ direction.
The angle $\theta_+$ is assigned to $l^+$ or $\pi-\theta_+$ to the $l^-$,
 as described above for single photon final state radiation. The
azimuthal angle of the total photon momentum $\vec{K_2}$ is then
generated uniformly by assigning the angle $\phi_{\gamma T}'$. The
3-axis is now rotated into the direction of the incoming $e^+$ and the
angle $\phi_+$ (or $\phi_-$) is assigned. Finally the 3-momentum
of the remaining lepton is found using momentum conservation in
the LAB system.

\SUBSUBSECTION{\tt{Three Final State Photons}}

Events with three final state photons are generated, according to the
obvious generalisation of Eqn.(4.66), by the subroutine ZFIN3G.
The event configuration is defined by the following kinematical
variables:

  \[ y,\mbox{\hspace{0.5cm}} \cos\theta_{\gamma 1}',\mbox{\hspace{0.5cm}}
  \cos\theta_{\gamma 2}',\mbox{\hspace{0.5cm}} \cos\theta_{\gamma 3}',
  \mbox{\hspace{0.5cm}} u',\mbox{\hspace{0.5cm}} u,
  \mbox{\hspace{0.5cm}} \cos\theta_{\pm},\mbox{\hspace{0.5cm}}
  \phi_{12}', \mbox{\hspace{0.5cm}} \phi_{\gamma3}',
  \mbox{\hspace{0.5cm}} \phi_{\gamma T}',
  \mbox{\hspace{0.5cm}} \phi_{\pm} \]

$\theta_{\gamma 3}'$ is similarly defined to $\theta_{\gamma 1}'$,
$\theta_{\gamma 2}'$ but refers to the third (least energetic) photon.
The variable $u'$ is given by:
\begin{equation} u' = \ln\left(\frac{y_1+y_2}{y_3}\right)
\end{equation}
The definition of $\phi_{\gamma 3}'$ is similar to that given above
for the case of 2 final state photons except that the total photon
momentum vector $\vec{K_3}$ corresponds to three photons. The angle
$\phi_{\gamma 3}'$ is that between the plane defined by the total
momentum of the first two (most energetic) photons $\vec{K_2}$
and the $l^+$ (or $l^-$), and that defined by the direction of
photon 3 and $l^+$ (or $l^-$). $l^+$ ($l^-$) is taken for
$\cos\theta_{\gamma 3}'  > 0 ( < 0 )$.
 All other variables have been defined
previously.

Following the soft photon limit of the generalisation of Eqn.(4.66) the
photon energies are first generated according to the distribution:
\begin{equation}
 \frac{d^3n}{dy_1dy_2dy_3} \simeq \frac{1}{y_1y_2y_3} \end{equation}
By a change of variables:
  \[ W=\ln[1/(y_1+y_2+y_3)],\mbox{\hspace{1.0cm}}
   u = \ln(y_1/y_2),\mbox{\hspace{1.0cm}}
   u' = \ln[(y_1+y_2)/y_3] \]

(4.74) simplifies to
\begin{equation}
 \frac{d^3n}{dWdudu'} \simeq constant \end{equation}
A 1-D LUT for the variable $W$, and a 2-D LUT for $u'$ (for a given $W$)
are created at initialisation by the subroutines SETYD3, SETRK3
respectively. The limits on the variable $u$ in the nested integration:

\[ I = \int dW \left\{\int du'\left [ \int du \right] \right\}  \]

are
\begin{eqnarray}
 u_{MIN} = \ln \left( \frac{y'-y_1^{MAX}}{y_1^{MAX}}\right) \\
 u_{MAX} = \ln \left( \frac{y'-y_1^{MIN}}{y_1^{MIN}}\right)
\end{eqnarray}
where
\begin{eqnarray*}
 y^{MIN}_1  = \max[y'-y_{MAX},y_{MIN}]  \\
 y^{MAX}_1  = \min[y'-y_{MIN},y_{MAX}]
\end{eqnarray*}
and
 \[ y'=y_1+y_2=y e^{u'}/(1+e^{u'}) \]

The limits on the $u'$ integration are given by Eqns.(4.50,4.51) with
the
replacements $u \rightarrow u'$ and
\begin{eqnarray*}
 y^{MIN}_1  \rightarrow \max[2y_{MIN},y-y_{MAX}]  \\
 y^{MAX}_1  \rightarrow \min[y_{MAX},y-2y_{MIN}]
\end{eqnarray*}
The integral over $u$ is done analytically, that over $u'$ by Gaussian
integration, and that over $W$ by Simpson's rule. During the event
generation phase the value of $W$ is chosen from the 1-D LUT, The value
of $u'$ is then chosen using the 2-D LUT. Finally, $u$ is given by:
\begin{equation} u = u_{MIN}+Rn(u_{MIN}-u_{MAX}) \end{equation}
where $u_{MIN}$, $u_{MAX}$ are given by Eqns.(4.76,4.77).
$y_1$, $y_2$, $y_3$ are then derived from $W$,$u$, $u'$ via the
equations:
\begin{eqnarray}
 y & = & e^W       \nonumber \\
 y' & = & ye^{u'}/(1+e^{u'}) \nonumber \\
 y_1& = & y'e^{u}/(1+e^{u}) \\
 y_2& = & y'-y  \nonumber  \\
 y_3& = & y -y' \nonumber
\end{eqnarray}
The photon spectra are then modified by weight rejection to take into
account the hard photon corrections given by the Gribov-Lipatov~\cite{x38}
kernels.

Defining the weight function:
\begin{equation} {\cal W} = d(y_1)d(y_2)d(y_3)y_1y_2y_3 \end{equation}
where $d(y)$ is defined after Eqn.(4.40), the event is rejected if:
 \[ {\cal W} < Rn \]
The construction of the momentum vectors of the final state particles
follows closely that described in the previous section for the case
of two photons. The 3-vectors of the two most energetic photons are
given by Eqns(4.68,4.69) above, and the angle $\theta_{\gamma 2}'$
is chosen relative to the $l^+$ direction (opposite of the $l^-$
direction ) according to whether $\cos \theta_{\gamma 2}'$ is $> 0$ or
($ < 0$). Energy conservation is then checked using Eqn.(4.54). To take
into account recoil effects in the radiation of the third photon
the acollinearity angle between  the $l^+$ and $l^-$ after the first
two photons have been radiated is calculated, and $\theta_{\gamma 3}'$
is chosen relative to the $l^+$ direction (opposite of the $l^-$
direction, for $\cos \theta_{\gamma 3}' > 0$ ($ < 0$). That is,
the procedure used above for the second photon is iterated. The
azimuthal angle $\phi_{\gamma 3}'$ is chosen to be that between
the planes defined by ($\vec{K_2},\vec{q_+}$) ; ($\vec{k_3},\vec{q_+}$)
for $\cos \theta_{\gamma 3}'  > 0$, and that between
($\vec{K_2},\vec{q_-}$) ; ($\vec{k_3},\vec{q_-}$)
for $\cos \theta_{\gamma 3}' < 0$. Since the relative directions of
all three photons are now fixed energy conservation is again checked
using the generalisation of Eqn.(4.53):
\begin{equation}
\sqrt{s}   >  k_1+k_2 +k_3+|\vec{k_1}+\vec{k_2}+\vec{k_3}|
\end{equation}
If the RHS of Eqn(4.81) is $ > 0.99\sqrt{s}$, the event is rejected.
Eqns.(4.71,4.72)
with the replacement $\vec{K_2} \rightarrow \vec{K_3}$ is
now used to calculate $y_+$, $y_-$. The scattering angle $\theta_+$
is assigned to the $l^+$, or $\pi-\theta_+$ to the $l^-$, according to
Eqns.(4.61). The angles $\phi_{\gamma T}'$ and $\phi_+$ or
$\phi_-$ are assigned as described above for the two photon case.
Finally the 3-momentum of the remaining lepton is calculated from
overall momentum conservation.

\SUBSUBSECTION{\tt{Events with One Initial State and One Final State
 Photon}}

Events with one initial state and one final state photon are
generated by the subroutine ZINF2G according to the differential
cross-section (c.f Eqns.(4.40,4.66)):
 \begin{eqnarray}
 d\sigma^IF_{2\gamma} & = & C_{n} d\sigma_{0}(s'')
 \frac{p_{+} \cdot p_{-}}{(p_{+} \cdot k_{1})
 (p_{-} \cdot k_{1})}\frac{q_{+} \cdot q_{-}}{(q_{+} \cdot k_{2})(
 q_{-} \cdot k_{2})}\nonumber \\
   &   & \mbox{} \times (1-\frac{k_{1}}{E}+\frac{k_{1}^{2}}{2E^{2}})
(1-\frac{k_{2}}{E}+\frac{k_{2}^{2}}{2E^{2}}) \frac{d^{3}k_{1}}{k_{1}}
\frac{d^{3}k_{2}}{k_{2}} \nonumber \\
  &   & \nonumber \\
  & = & 4C_{n}\frac{d\sigma_{0}(s'') d(y_{1}) d(y_{2}) dy_{1}
d\Omega_{1} dy_{2}d\Omega_{2}}{(1-\beta_{IN}^{2}\cos^{2}\Theta_
{\gamma 1})(1-\beta_{FI}^{2}\cos^{2}\Theta_{\gamma 2}')}
\end{eqnarray}
where
\[  s'' = s(1-y_1) \]
and photons 1,(2) are radiated in the initial,(final) state.
The kinematical variables used, in this case, to define the event
configuration are:

  \[ y,\mbox{\hspace{0.5cm}} cos\theta_{\gamma 1} ,\mbox{\hspace{0.5cm}}
  cos\theta_{\gamma 2}',\mbox{\hspace{0.5cm}} u,
  \mbox{\hspace{0.5cm}}cos\theta_+'',\mbox{\hspace{0.5cm}}
  \phi_+'', \mbox{\hspace{0.5cm}} \phi_{\gamma 2}'',
  \mbox{\hspace{0.5cm}} \phi_{\gamma 1} \]

Here $\theta_{\gamma 2}''$ is the angle between photon 2 and the $l^+$
in the rest frame of $l^+l^-\gamma 2$ (ODLGR frame). In this frame the
1-axis is chosen perpendicular to the plane defined by the incoming
$e^+$ and the direction of photon 1. $\theta_+''$, $\phi_+''$ are
the polar and azimuthal angles of the $l^+$ relative to the
incoming $e^+$ direction (allowing, if necessary, for the recoil
generated by photon 1) in the ODLGR frame. The other variables
have been previously defined.

The variables $y$, $u$ are generated using 1-D, 2-D  LUT created at
initialisation by the subroutines SETYDI, SETR respectively.
The procedure is the same as that described in Section 4.3.2 above.
The 2-D LUT is created according to Eqn.(4.49) with the replacement:
$\tilde{s}' \rightarrow s''$. The configuration of $l^+l^- \gamma 2$
is generated as described in Section 4.4.1, for single final state
radiation, but in the ODLGR frame rather than the LAB frame. Recoil
effects from the initial state photon are accounted for by rotation by
the angle $\alpha_B$ given by Eqns.(4.34-4.35) with the replacement
$y \rightarrow y_1$ in the formula (4.36) for $\beta^{\star}$. Since
the scaled photon energies $y_1$, $y_2$ are defined in the LAB frame,
the energy of photon 2 must first be calculated, by Lorentz
transformation, in the ODLGR frame, from its angles in this frame
(specified by $\theta_{\gamma 2}''$, $\phi_{\gamma 2}''$,
$\theta_+''$, $\phi_+''$ ) and its LAB energy $\frac{\sqrt{s}}{2} y_2$.
For simplicity, in this case, $\theta_+''$ is always assigned to the
$l^+$, rather than $\theta_+''$ to the $l^+$ or $\pi-\theta_+''$ to the
$l^-$ according to Eqn(4.61) \footnote{The procedures described in
Ref.[14] for assigning the scattering angle in the case of single
initial or final state photons are not directly applicable in this case}.
Finally the 4-vectors of $l^+$, $l^-$ and photon 2 are tranformed back into
the lab. frame and the event is rotated about the incoming $e^+$ direction
so that photon 1 has azimuthal angle $\phi_{\gamma 1}$.
\SECTION{\bf{Program Structure and Performance}}
\SUBSECTION{General Organisation. How to use the Program}
The program has a very short main program containing definitions of the most
important input parameters, which are stored in the labelled common block ICOM.
These variables are described in Table 3. The execution of the program has
three distinct phases:
\begin{itemize}
\item Initialisation
\item Generation of a single unit weight event
\item Termination
\end{itemize}
Each of these phases is entered via a call to subroutine BHAGENE3 in the main
program:
\\[.5cm]
CALL BHAGENE3(MODE,CTP1,CTP2,CTM1,CTM2,CTAC,EP0,EM0)
\\[.5cm]
 MODE is set to $-1,0,1$ for the initialisation, generation and termination
 phases respectively. The other parameters of BHAGENE3 define the kinematical
 cuts to be applied to the generated events:
 \\[.5cm]
\hspace*{2cm} CTP1 $=$ minimum value of $\cos \theta_{l^+}$
\\
\hspace*{2cm} CTP2 $=$ maximum value of $\cos \theta_{l^+}$
\\
\hspace*{2cm} CTM1 $=$ minimum value of $\cos \theta_{l^-}$
\\
\hspace*{2cm} CTM2 $=$ maximum value of $\cos \theta_{l^-}$
\\
\hspace*{2cm} CTAC $=$ maximum value of $\cos \phi_{col}$
\\
\hspace*{2cm} EP0 $=$ minimum energy of $l^+$ ($GeV$)
\\
\hspace*{2cm} EM0 $=$ minimum energy of $l^-$ ($GeV$)
 \\
\\[.5cm]
 All these cuts are applied in the laboratory (incoming $e^+,e^-$ centre of
 mass) system. The angle $\phi_{col}$ is the collinearity angle between the
 $l^+$ and the $l^-$ (CATC $=$ -1 for a back-to-back configuration).
 In the calls of BHAGENE3 with MODE $=$ 0,1 only this parameter need be
 specified.
  A typical main program to generate 5000 unit weight events might be:
\\[.5cm]
\\
 PROGRAM BHAMAIN
\\
 IMPLICIT REAL*8  (A-H,O-Z))
\\
 COMMON/ICOM/OIMZ,OIMT,OIMH,OMAS,IOCHA,IOEXP,OW,OCTC1,OCTC2,IOXI
\\
 OIMZ=91.18D0~~~~~~~~~~~~~~~~~~!  {\sf Z Mass (GeV)}
\\
 OIMT=150.0D0~~~~~~~~~~~~~~~~~~!  {\sf Top quark mass (GeV)}
\\
 OIMH=100.0D0~~~~~~~~~~~~~~~~~~!  {\sf Higgs boson mass (GeV)}
\\
 OMAS=0.12D0~~~~~~~~~~~~~~~~~~~!   {\sf Alphas }
\\
 IOCHA=1~~~~~~~~~~~~~~~~~~~~~~~~~!  {\sf 0 for Muon pairs, 1 for
electron pairs }
\\
 IOEXP=1~~~~~~~~~~~~~~~~~~~~~~~~~!   {\sf 0 for O(Alpha), 1 for
exponentiation }
\\
 OW=91.00D0~~~~~~~~~~~~~~~~~~~~~!  {\sf Collision energy (GeV) }
\\
 OCTC1=-0.8D0~~~~~~~~~~~~~~~~~~!  {\sf Lower cos(theta) limit in ODLR frame }
\\
 OCTC2=0.8D0~~~~~~~~~~~~~~~~~~~!  {\sf Upper cos(theta) limit in ODLR frame }
\\
 IOEXI=713883717~~~~~~~~~~~~~~~!  {\sf Intial random number }
\\
\\
\hspace*{3cm} ! {\sc KINEMATICAL CUTS FOR GENERATED EVENTS }
\\
\\
 CTP1=-0.7D0~~~~~~~~~~~~~~~~~~~~!  {\sf Lower cos(theta) for l+ in the
lab frame}
\\
 CTP2=0.7D0~~~~~~~~~~~~~~~~~~~~~!  {\sf Upper cos(theta) for l+ in the
lab frame}
\\
 CTM1=-0.7D0~~~~~~~~~~~~~~~~~~~~!  {\sf Lower cos(theta) for l- in the lab
frame}
\\
 CTM2=0.7D0~~~~~~~~~~~~~~~~~~~~~!  {\sf Upper cos(theta) for l- in the lab
frame}
\\
 CTAC=-0.9D0~~~~~~~~~~~~~~~~~~~~!  {\sf Upper cosine of collinearity angle }
\\
 EP0=2.0D0~~~~~~~~~~~~~~~~~~~~~~~~!  {\sf Minimum energy of l+ (GeV) }
\\
 EM0=2.0D0~~~~~~~~~~~~~~~~~~~~~~~!  {\sf Minimum energy of l- (GeV) }
\\
\\
 \hspace*{3cm} !  {\sc INITIALISATION PHASE }
\\
\\
 CALL BHAGENE3(-1,CTP1,CTP2,CTM1,CTM2,CTAC,EP0,EM0)
\\
\\
\hspace*{3cm} !   {\sc GENERATION LOOP }
\\
\\
 DO J=1,5000
\\
 CALL BHAGENE3(0)
\\
 ENDDO
\\
\\
 \hspace*{3cm}  !  {\sc TERMINATION PHASE }
\\
\\
 CALL BHAGENE3(1)
\\
 STOP
\\
 END
\\
\\
Other initialisation parameters of interest to users are defined in
BHAGENE3 itself. A list of the most important of these can be found in Table 4.
\SUBSECTION{Initialisation Phase}
A flow chart of the initialisation phase of the program is shown in Fig 1.
The functionality of the different subprograms shown there has been described
above. The main physical quantities calculated are also indicated. In order to
estimate the average weights of events with $\geq$ two hard photons the
initialisation phase actually includes the generation of $4\times 10^{4}$
events,
which is relatively time consuming. Users should be aware of this.
\SUBSECTION{Generation Phase}
As shown in the flow chart in Fig 2. each of the different event topologies is
generated by a different subprogram. The calculations performed by the five
different
hard photon subgenerators : ZINIGB, ZFINGB, ZFIN2G, ZFIN3G, and ZINF2G
 are described above
in Section 4. The 4-vectors of the generated events are written in the
labelled
common
block C4VEC :
\\[.5cm]
\hspace{2cm} COMMON/C4VEC/PPV(4),PMV(4),
 QPV(4),QMV(4),GAM1V(4),GAM2V(4),GAM3V(4),\newline
WEIGHT,NPHOT,ISG
\\[.5cm]
The 4-vectors (defined as: $p_x,p_y,p_z,E$ ) are in the order:
($e^+,e^-,l^+,l^-,
\gamma_1,\gamma_2,\gamma_3 $ ). WEIGHT is the event weight, NPHOT the number
of
filled
photon 4-vectors and ISG a flag indicating the subgenerator that produced
the event
(see Table 2.). The photons are ordered in energy, the first photon being the
most
energetic.
\SUBSECTION{Termination Phase}
A flow chart of the termination phase is shown in Fig 3. The exact
cross-section
($\sigma^{CUT}$) and its error ($\Delta\sigma^{CUT}$) are printed out,
together
with
the input parameters (including those calculated in the initialisation phase).
Other
cross-sections used to calculate the event generation probabilities
 $P(n_{\gamma}^I,n_{\gamma}^F)$ are also printed out.
  A sample output is shown in Fig 4.
\SUBSECTION{Program Performance}
As mentioned above, after a relatively lengthy initialisation procedure, during
which all look-up tables are created and average weights are calculated for
multiphoton events, the event generation procedure is itself fast.
Typical times for the initialisation phase are 140, 48 IBM 3090 CPU seconds for
$e^+e^-$, $\mu^+\mu^-$ pair generation, respectively. Average times to generate
 a single unit weight $e^+e^-$, ($\mu^+\mu^-$) event are $1.41\times 10^{-3}$,
($0.64\times 10^{-3}$) IBM 3090 CPU seconds. The combined weight distribution
for initial and final state radiation $e^+e^-$ events for parameters and cuts
as
 in the sample main program given above, (whose output is shown in Fig. 4) is
 presented in Fig. 5.  The weight distribution can be seen to be well centered
 around 1.0, resulting in an efficient generation of unit weight events by the
 weight throwing procedure ~\cite{x4}. The fractions of events with weights
  $>$ 2.0, (3.0) is $\simeq 6 \times 10^{-3}$, ($ 2 \times 10^{-4}$)
  respectively. The maximum weight may then be chosen to be 2.0, allowing
  efficient generation of event samples with cross-sections known at, or below
  the \% level~\cite{x10}.

\pagebreak
{\bf \large {Appendix A}}

Following Ref.[39], the Born differential cross-section for Bhabha
scattering, including both $s$ and $t$ channel Z exchange may be written
as:
\[
\frac{d\sigma_{0}}{dt} = \frac{2 \pi \alpha^2}{s^2}[B_0+B_2(\frac{t}{s})
^2+B_3(1+\frac{t}{s})^2] \hspace{2cm}  (A1) \]
 where
\begin{eqnarray*}
      t & = & -\frac{s}{2}(1-\cos\theta_+)     \\
    B_0 & = &  [\frac{s}{t}+c_- \chi_t(t)]^2   \\
    B_2 & = &  |1+c_- \chi_s(s)|^2
\end{eqnarray*}
\[ B_3 = \frac{1}{2}\{|1+\frac{s}{t}+a_+[\chi_t(t)+\chi_s(s)]|^2+
          (a_+ \rightarrow a_-) \} \]
\begin{eqnarray*}
    c_- = g_V^2-g_A^2    & , &  a_{\pm} = (g_V \pm g_A)^2  \\
    \chi_t(t) = \frac{s}{t-M_Z^2} & , &  \chi_s(s) = \frac{s}{s-M_Z^2
 +is(\Gamma_Z/M_Z)}
\end{eqnarray*}

$\theta_+$ is the scattering angle between the incoming and outgoing
$e^{\pm}$. $M_Z$, $\Gamma_Z$ are the mass and width of the Z. The Z width
is neglected in the t-channel exchange amplitude. The vector and
axial-vector coupling constants $g_A$, $g_V$ are given, in the Standard
Model by Eqns. 3.3-3.6. The cross-section for $\mu$-pair production is
recovered on setting $s/t$ to zero in $B_0$, $B_3$.

Analytical integration of $A$1 over $t$ yields the result ~\cite{x34}
( see also the first of Ref.[4]):
\[
\int_{t_{MIN}}^{t_{MAX}} d \sigma_0 = \frac{2 \pi \alpha^2}{s^2}
[S(t_{MIN})-S(t_{MAX})] \hspace{2cm} (A2) \]

The function $S(t)$ is the sum of the following 5 terms derived from
Eqn. $A$1 :
\begin{eqnarray*}
\int B_0 dt &=& s^2 [-\frac{1}{t} + \frac{2c_-}{M_Z^2} \ln
 \frac{t-M_Z^2}{t} - \frac{c_-^2}{t-M_Z^2}] \\
  & & \\
 \int B_3 dt &=& \frac{1}{2}\{ s^2[ -\frac{2}{t} + \frac{2(a_+ +a_-)}
{M_Z^2} \ln \frac{t-M_Z^2}{t}-\frac{(a_+^2+a_-^2)}{t-M_Z^2}]   \\
 & & +2s[(b_++b_-) \ln t + (a_+b_+ +a_-b_-) \ln (t-M_Z^2)] + (b'_+
+ b'_-)t \}   \\
  & & \\
 \int \frac{2 B_3 t}{s} dt &= & s\{2 \ln t + 2 (a_+ + a_-) \ln (t-M_Z^2)+
(a_+^2 + a_-^2)[\ln(t-M_Z^2)- \frac{M_Z^2}{t-M_Z^2}]\} \\
& &  + 2\{(b_+ + b_-)t +(a_+ b_+ + a_-b_-)[t+M_Z^2\ln(t-{M_Z^2}]\}
  +(b'_+ + b'_-) \frac{t^2}{2s} \\
  & & \\
\int B_2 \frac{t^2}{s^2} dt &=  &\frac{B_2 t^3}{3 s^2}  \\
  & & \\
\int B_3 \frac{t^2}{s^2} dt &=& \frac{1}{2}\{[2t+2(a_+ + a_-)(t+M_Z^2\ln(t-M_Z^
2))+(a_+^2 +a_-^2)(t+2M_Z^2 \ln (t-M_Z^2)   \\
& & -\frac{M_Z^4}{t-M_Z^2})] + \frac{2}{s}[(b_+ + b_-)
 \frac{t^2}{2} + (a_+ b_+ + a_- b_-)
(\frac{t^2}{2}+t M_Z^2+M_Z^4\ln(t-M_Z^2))] \\
& & +(b'_+ +b'_-)\frac{t^3}{3s^2}\}
\end{eqnarray*}
where
\begin{eqnarray*}
b_{\pm} &=& 1 + a_{\pm} Re\chi_s  \\
b'_{\pm} &=& 1 + a_{\pm}2 Re\chi_s + a_{\pm}^2|\chi_s|^2
\end{eqnarray*}


{\bf \large {Appendix B}}

The exact hard photon cross-section, with exponentiated initial state
radiation, is derived from formulae given in the second of Refs.[4] :

\[  \frac{d\sigma^{EXACT}}{d\Omega_+ d\Omega_{\gamma}dy} =
\frac{\alpha^3 y}{16\pi^2 s} X(\alpha_k) \hspace{2cm} (B1) \]

The function $X$($\alpha_k$) of the kinematical variables $\alpha_k$
specifying the event configuration, (defined in Sections 4.3.1 to
4.4.4), has contributions from $s$-channel exchanges ($ss$), $t$-channel
exchanges ($tt$) and $s-t$ interference ($st$). In each of these cases
separate
contributions from initial state radiation (INI), final state
radiation (FIN), and initial/final interference (INT) may be
distinguished. Thus :

\[ X=\sum_{i,j} X_i^j \hspace{.5cm} i=ss,tt,st;\hspace{.5cm}
j=INI,FIN,INT \hspace{1cm} (B2) \]

where

\begin{eqnarray*}
X_{ss}^{INI} & = & \frac{1}{s'\kappa_+\kappa_-}\left[B_1(c_-,s')
[f(s)t^2+t'^{2}]+B_5(s')[f(s)u^2+u'^{2}]\right]  \\
  &  & -m_e^2f(s) \left[\frac{B_{ss}(s',t,u)}{\kappa_-^2} +
\frac{B_{ss}(s',t',u')}{\kappa_+^2}\right]   \\
  &  & \\
X_{ss}^{FIN} & = & \frac{1}{s\kappa'_+\kappa'_-}\left[B_1(c_-,s)
[t^2+t'^{2}]+B_5(s)[u^2+u'^{2}]\right]  \\
  &  & -m_l^2 \left[\frac{B_{ss}(s,t,u')}{\kappa_{-}^{'2}} +
\frac{B_{ss}(s,t',u)}{\kappa_{+}^{'2}}\right]   \\
  &  & \\
X_{ss}^{INT} & = & \frac{1}{ss'}\left[\frac{u}{\kappa_+\kappa'_-}
+\frac{u'}{\kappa_-\kappa'_+}-\frac{t}{\kappa_+ \kappa'_+}
-\frac{t'}{\kappa_- \kappa'_-} \right] \\
  &  & \times \left[B_4(s,s')(t^2+t'^2)+B_6(s,s')(u^2+u'^2) \right] \\
  &  & +\frac{\vec{p_+}\cdot(\vec{q_+}\times \vec{q_-})(s-s')}
{2 \sqrt{s} s' \kappa_+ \kappa_- \kappa'_+ \kappa'_-}
B_7(s,s')(u^2-u'^2) \\
  &  & \\
X_{tt}^{INI} & = & \frac{s}{\kappa_+\kappa_-}\left[ B_6(t,t')
[\frac{f(s)u^2+u'^2}{tt'}]+B_4(t,t')[\frac{f(s)s^2+s'^2}{tt'}]\right] \\
  &  & -m_e^2f(s) \left[\frac{B_{tt}(s',t,u)}{\kappa_-^2} +
\frac{B_{tt}(s',t',u')}{\kappa_+^2}\right]   \\
  &  & \\
X_{tt}^{FIN} & = & \frac{s'}{\kappa'_+\kappa'_-}\left[ B_6(t,t')
[\frac{u^2+u'^2}{tt'}]+B_4(t,t')[\frac{s^2+s'^2}{tt'}]\right]\\
  &  & -m_l^2 \left[\frac{B_{tt}(s,t,u')}{\kappa_-^{'2}} +
\frac{B_{tt}(s,t',u)}{\kappa_+^{'2}}\right]   \\
  &  & \\
X_{tt}^{INT} & = & \frac{s^2+s'^2}{tt'} \left[-\frac{t}{\kappa_+
\kappa'_+}B_1(c_-,t')-\frac{t'}{\kappa_- \kappa'_-}B_(c_-,t)\right] \\
  &  & + \frac{u^2+u'^2}{tt'} \left[-\frac{t}{\kappa_+ \kappa'_+}
B_5(t')-\frac{t'}{\kappa_- \kappa'_-}B_5(t)\right] \\
  &  & + \left(\frac{u}{\kappa_+ \kappa'_-}+
\frac{u'}{\kappa_- \kappa'_+}\right)
\left[ B_6(t,t')
[\frac{u^2+u'^2}{tt'}]+B_4(t,t')[\frac{s^2+s'^2}{tt'}]\right]
\\  &  & \\
X_{st}^{INI} & = & \frac{f(s)u^2+u'^2}{s'\kappa_+ \kappa_-}\left[
-\left(\frac{u'+t'}{t'}\right)B_6(s',t')-\left(\frac{u+t}{t}\right)
B_6(s',t)\right] \\
  &  & -m_e^2f(s)\left[\frac{B_{st}(s',t,u)}{\kappa_-^2}
 +\frac{B_{st}(s',t',u')}{\kappa_+^2}\right] \\
  &  & \\
X_{st}^{FIN} & = & \frac{u^2+u'^2}{s\kappa'_+ \kappa'_-}\left[
-\left(\frac{u'+t}{t}\right)B_6(s,t)-\left(\frac{u+t'}{t'}\right)
B_6(s,t')\right] \\
  &  & -m_l^2f(s)\left[\frac{B_{st}(s,t,u')}{\kappa_-^{'2}}
 +\frac{B_{st}(s,t',u)}{\kappa_+^{'2}}\right]     \\
  &  &   \\
\end{eqnarray*}
\begin{eqnarray}
X_{st}^{INT} & = & \left[\frac{u'}{\kappa_- \kappa'_+}+
\frac{u'+s'}{\kappa_+\kappa'_+}\right]\left(\frac{u^2+u'^2}{s't'}\right)
B_6(s',t') \nonumber \\
  &   & +\left[\frac{u}{\kappa_+ \kappa'_-}+
\frac{u+s'}{\kappa_- \kappa'_-}\right]\left(\frac{u^2+u'^2}{s't}\right)
B_6(s',t) \nonumber \\
  &   & +\left[\frac{u'}{\kappa'_+ \kappa_-}+
\frac{u'+s}{\kappa'_- \kappa_-}\right]\left(\frac{u^2+u'^2}{st}\right)
B_6(s,t) \nonumber \\
  &   & +\left[\frac{u}{\kappa_+ \kappa'_-}+
\frac{u+s}{\kappa'_+\kappa_+}\right]\left(\frac{u^2+u'^2}{st'}\right)
B_6(s,t') \nonumber \\
  &   & +\frac{\sqrt{s}[\vec{p_+}\cdot(\vec{q_+}\times\vec{q_-})]
(u^2-u'^2)}{2\kappa_+\kappa_-\kappa'_+\kappa'_-}
[-\left(\frac{t-t'}{tt'}\right)B_7(t,t')
+2\frac{B_7(s,t)}{st(\kappa_-\kappa'_+\kappa'_-)} \nonumber \\
  &   & +2\frac{B_7(s,t')}{st'(\kappa_+\kappa'_+\kappa'_-)}
 -2\frac{B_7(s',t)}{s't(\kappa_+\kappa_-\kappa'_-)}
 -2\frac{B_7(s',t')}{s't'(\kappa_+\kappa_-\kappa'_+)}] \nonumber
\end{eqnarray}
\begin{eqnarray*}
B_1(c,s) & = & 1+[\frac{2c}{s}(s-M_Z^2)+c^2]|\chi_s(s)|^2 \\
  &  & \\
B_4(s,s') & = & 1+c_-[(1-\frac{M_Z^2}{s})|\chi_s(s)|^2
+(1-\frac{M_Z^2}{s'})|\chi_s(s')|^2]  \\
&  & +c_-^2[(1-\frac{M_Z^2}{s})(1-\frac{M_Z^2}{s'})
+\frac{M_Z^2 \Gamma_Z^2}{ss'}]|\chi_s(s)|^2|\chi_s(s')|^2 \\
  &  & \\
B_5(s) & = & 1+[2(g_V^2+g_A^2)(1-\frac{M_Z^2}{s})
+(g_V^4+g_A^4+6g_V^2g_A^2)]|\chi_s(s)|^2
\end{eqnarray*}
\begin{eqnarray*}
B_6(s,s') & = & 1+(g_V^2+g_A^2)\left[(1-\frac{M_Z^2}{s})|\chi_s(s)|^2+
(1+\frac{M_Z^2}{s'})|\chi_s(s')|^2 \right] \\
  & = & +(g_V^4+g_A^4+6g_V^2g_A^2)\left[(1-\frac{M_Z^2}{s})
(1-\frac{M_Z^2}{s'})+\frac{M_Z^2 \Gamma_Z^2}{ss'}\right]
|\chi_s(s)|^2 |\chi_s(s')|^2
\\  &  & \\
B_7(s,s') & = & -4M_Z \Gamma_Z \left\{ g_A g_V [\frac{|\chi_s(s)|^2}{s}-
\frac{|\chi_s(s')|^2}{s'} ]
\right.
\\
  &   &
\left.
+2 g_V g_A(g_V^2+g_A^2)(\frac{1}{s}-\frac{1}{s'})
|\chi_s(s)|^2 |\chi_s(s')|^2
\right\}
\\  &  & \\
B_{ss} (s,t,u) & = & \left[ B_{1}(a_{+}^{2},s)+B_{1}(a_{-}^{2},s) \right]
(\frac{u}{s})^2+2B_{1}(c_{-},s)(\frac{t}{s})^2
\\  &  & \\
B_{tt}(s,t,u) & = & \left[ B_1(a_+^2,t)+B_1(a_-^2,t) \right]
(\frac{u}{t})^2+2B_1(c_-,t)(\frac{s}{t})^2
\\  &  & \\
B_{st}(s,t,u) & = & 4 B_6(s,t) \frac{u^2}{st}
\end{eqnarray*}

 The constants $c_-$, $a_{\pm}$ are defined in Appendix A and $g_V$,
$g_A$ in Eqns 2.3-2.6. Note that the functions $\chi_s$, $\chi_t$ are
assigned according to the arguments of $B_1$, $B_4$ ... . For example
$B_4(s,t')$ is given by the replacement $\chi_s(s') \rightarrow
\chi_t(t')$ in the expresssion above for $B_4(s,s')$.

 Exponentiation of the initial state radiation is included via the
function:
 \[ f(s) = 2 C_V^i \exp \beta_e \ln (y) - 1 \hspace{2cm} (B3) \]
where $\beta_e$, $C_V^i$ are defined after Eqn 2.1. This ensures that
the $y \rightarrow 0$ limit of Eqn. $B$1 is identical to the derivative
of Eqn 2.1 with respect to $y_0$, in which the replacement
$y_0 \rightarrow y$ is made. That is `soft' and `hard' photons are
treated in a consistent way.

{\bf \large {Appendix C}}

Two dimensional look up tables are generated according to a
generalisation of the procedure described in the text (Eqns 3.7-3.10)
for a one dimensional look up table. Consider for example the case of
the Born differential cross-section $d\sigma_0/dc$ for a range of
different CM energies : $ s_{min} < s < s_{max} $. Let $k$ be the bin
index for $s$. The equations 2.7, 2.8 generalise to:
\[ \sigma_0^k = \int_{c_{min}}^{c_{max}} \frac{d\sigma_0^k}{dc}dc
\hspace{2cm} (C1) \]

where

\[ d \sigma_0^k \equiv d \sigma_0 (s_k)   \]

\[ P_i^k =\frac{1}{\sigma_0^k}\int_{c_{min}}^{c_i} \frac{d \sigma_0^k}
{dc}dc \hspace{2cm} (C2) \]
For each value of $k$ the distribution $P_i^k$ is inverted by linear
interpolation to yield a look up table with bin index $j$ :
\[ c_i^j = f^k(P_i^k) \hspace{3cm} (C3) \]

To generate the angular distribution for a given value of $s$, the
adjacent bins in $s$ of index $k$, $k+1$ are first located :
\[ s_k < s < s_{k+1}\]
The closest bins in the look up tables of index $k$, $k+1$ to the
random number $Rn$ are the found:
\[P_i^k < Rn < P_{i+1}^k \]
\[P_{i'}^{k+1} < Rn < P_{i'+1}^{k+1} \]
With:
\[ \delta_i  = (Rn-P_i^k)/(P_{i+1}^k-P_i^k) \]
\[ \delta_{i'}=(Rn-P_{i'}^{k+1})/(P_{i'+1}^{k+1}-P_{i'}^{k+1}) \]
Two values of $c$ are now calculated according to :
\[ c^k = f^k(P_i^k)+[f^k(P_{i+1}^k)-f^k(P_i^k)]\delta_i \]
\[ c^{k+1} = f^{k+1}(P_{i'}^k)+[f^{k+1}(P_{i'+1}^{k+1})-f^{k+1}(P_{i'}^{k+1})]
\delta_{i'} \]
with
\[ \Delta_k = (s-s_k)/(s_{k+1}-s_k)  \]
the generated value of $c$ is, finally, given by the equation :
\[ c = c^k + (c^{k+1}-c^k) \Delta_k  \hspace{2cm} (C4) \]

\pagebreak

\pagebreak
\begin{table}
\begin{center}
\newpage
\begin{tabular}{|c|c|c|}
\hline
\hline
\multicolumn{1}{||c|}{ $n_{\gamma}^I$ } &
\multicolumn{1}{c|}{ $n_{\gamma}^F$ } &
\multicolumn{1}{c||}{ Assigned Weight  } \\
\hline
\hline
 2      & 0           & $\overline{W}'_I$    \\
 0      & 2           & $\overline{W}'_F$    \\
 0      & 3           & $\overline{W}'_I$   \\
 1      & 1           & $\sqrt{\overline{W}'_I \overline{W}'_F}$   \\
\hline
\end{tabular}
\caption[]{Weights assigned to multiphoton events }
\end{center}
\end{table}

\begin{table}
\begin{center}
\begin{tabular}{|c|c|c|c|}
\hline
\hline
\multicolumn{1}{||c|}{ $n_{\gamma}^I$ } &
\multicolumn{1}{c|}{ $n_{\gamma}^F$ } &
\multicolumn{1}{c|}{  P( $n_{\gamma}^I$ , $n_{\gamma}^F$ )  } &
\multicolumn{1}{|c||}{ ISG } \\
\hline
\hline
 0      & 0           & $P_{V,S}$ & 0   \\
 1      & 0           & $P_Ie^{-r_e}$ & 1   \\
 0      & 1           & $P_Fe^{-r_f}$ & 2   \\
 2      & 0           & $P_I(1-e^{-r_e})\rho_{IF}$ & 3   \\
 0      & 2           & $P_Fr_fe^{-r_f})\rho_{IF}$ & 4   \\
 0      & 3           & $P_F[1-(1+r_f)e^{-r_f}]\rho_{IF}$ & 6   \\
 1 & 1 & $[P_I P_F (1-e^{-r_e})(1-e^{-r_f}) ]^{\frac{1}{2}}\rho_{IF}$ & 5 \\
\hline
\end{tabular}
\caption[]{`A priori' probabilities for different hard photon
multiplicities }
\end{center}
\end{table}
\begin{table}
\begin{center}
\begin{tabular}{|c|c|}
\hline
 OIMZ       &  Z mass (GeV)   \\
 OIMT      & Top quark mass (GeV)   \\
 OIMH      & Higgs boson mass (GeV)   \\
 OMAS       & $\alpha_s(M_Z)$          \\
 IOCH      &   $=0  (\mu^+\mu^-) , =1 (e^+ e^-) $  \\
 IOEXP      &  $=1$ exponentiated , $=0$ $O(\alpha)$    \\
 OW      & collision energy (GeV)  \\
 OCTC1       &  lower $\cos \theta_{l^+}$ in the ODLR frame  \\
 OCTC2       &  lower $\cos \theta_{l^-}$ in the ODLR frame  \\
 IOXI      &  initial random number  \\
\hline
\end{tabular}
\caption[]{ Variables of the labelled common ICOM. OCTC1,OCTC2 are used in
setting up
the
LUT of the lepton scattering angles. To allow for the effects of the Lorentz
boost
the angular range should be chosen somewhat wider than that defined by the
cuts
in the LAB
system.}
\end{center}
\end{table}

\begin{table}
\begin{center}
\begin{tabular}{|c|c|}
\hline
 NPAR(1)       & \underline{1}, 0 weak loop corrections ON, OFF   \\
 NPAR(2)      & 2,\underline{3} parameterisations of had. vac. pol.   \\
 NPAR(3)    &  \underline{0},1,2 two-loop $ \alpha \alpha_s m_t^2 $
correction  \\
 NPAR(4)       & \underline{1},0 weak box diagrams ON, OFF       \\
 NPAR(6)      &   \underline{1}, 0  two-loop terms $\propto m_t^4$  ON,OFF  \\
 XPAR(1)      &   initial lepton charge (-1.D0)    \\
 XPAR(2)      &   final lepton charge (-1.D0)    \\
 XPAR(3)      &   final lepton colour (1.D0)    \\
 XPAR(4)       &  final lepton mass (GeV)     \\
 XPAR(9)       &  QCD correction to $\Gamma^Z_q$ (non-$b$ quarks)    \\
 XPAR(10)       &  QCD correction to $\Gamma^Z_b$    \\
 YMA       &  maximum value of $\sum E_{\gamma}/E_{beam}$  (0.99D0)   \\
 YMI       &  minimum value of $ E_{\gamma}/E_{beam}$  (0.005D0)   \\
 WTMAX       &  maximum value of the event weight (2.0D)  \\
\hline
\end{tabular}
\caption[]{ Control parameters defined in SUBROUTINE BHAGENE3. Default values
are underlined or given
in parentheses.}
\end{center}
\end{table}

\clearpage
\begin{center}
{\bf FIGURE CAPTIONS}
\end{center}
\begin{itemize}
\item[{\bf Fig 1}] Flow Chart of the Initialisation Phase.
\item[{\bf Fig 2}] Flow Chart of the Event Generation Phase.
\item[{\bf Fig 3}] Flow Chart of the Termination Phase.
\item[{\bf Fig 4}] A typical line-printer output from BHAGENE3.
\item[{\bf Fig 5}] Distribution of weights for events with hard
initial or final state photons.
\end{itemize}
\end{document}